\documentclass[11pt]{article}
\usepackage{a4}
\usepackage{amsmath}
\usepackage{amssymb}
\usepackage[dvips]{graphicx}
\usepackage{color}
\def\beq{\begin{equation}}
\def\eeq{\end{equation}}
\def\beqa{\begin{eqnarray}}
\def\eeqa{\end{eqnarray}}
\def\be{\begin{align}}

\def\n{\nonumber \\}

\begin{document}

\begin{flushright}
SAGA-HE-282\\
KEK-TH-1771
\end{flushright}
\vskip 0.5 truecm

\begin{center}
{\Large{\bf 
Evolution of Vacuum Fluctuations \\
 of an Ultra-Light Massive Scalar Field \\
 generated during and before Inflation 
}}\\
\vskip 1cm

{\large Hajime Aoki$^{a}$ and 
 Satoshi Iso$^{b,c}$ 
}
\vskip 0.5cm

$^a${\it Department of Physics, Saga University, Saga 840-8502,
Japan  }\\
$^b${\it KEK Theory Center, 
High Energy Accelerator Research Organization (KEK),
Ibaraki 305-0801, Japan }\\
$^c$
{\it  Graduate University for Advanced Studies (SOKENDAI), \\
Ibaraki 305-0801, Japan}
\end{center}

\vskip 1cm
\begin{center}
\begin{bf}
Abstract
\end{bf}
\end{center}

We consider an ultra-light scalar field with a mass comparable to (or lighter than) 
the Hubble parameter of the present universe, 
and calculate the time evolution of the energy-momentum 
tensor of the vacuum fluctuations generated during and before inflation until the 
late-time radiation-dominated and matter-dominated universe. 
The equation of state changes from $w=1/3$ in the early universe to $w=-1$ at present,
and it can give a candidate for the dark energy that we observe today.
It then oscillates between $w=-1$ and $1$ with the amplitude of the energy density decaying as $a^{-3}$.
If the fluctuations are generated during ordinary inflation with the Hubble parameter 
$H_I \lesssim 10^{-5} M_{\rm Pl}$, where $M_{\rm Pl}$ is the reduced Planck scale,
we need a very large e-folding number $N \gtrsim 10^{12}$
to explain the present dark energy of the order of $10^{-3} {\rm eV}$. 
If a Planckian universe with a large Hubble parameter $H_P \sim M_{\rm Pl}$ existed 
 before the ordinary inflation, an e-folding number $N \sim 240$ of the Planckian inflation is sufficient.

\newpage
%%%%%%%%%%%%%%%%%%
\section{Introduction}
\label{sec:intro}
\setcounter{footnote}{0}
\setcounter{equation}{0}

Our universe is well  described by the 
spatially flat $\Lambda$ cold dark matter ($\Lambda$CDM) model. 
According to the PLANCK 2013 results~\cite{PlanckCosm}, 
only 5.1\% of the energy density is attributed to
a known form of baryonic matter, while 26.8\% 
is attributed to cold dark matter, and 68.3\% to
 dark energy (DE).
Although its equation 
of state $w(=p/\rho)=-1$ seems like that of vacuum energy 
of quantum fields, 
there is no reasonable explanation for its magnitude,
$\rho_{\rm{DE}}= 3 (M_{\rm Pl} H_0)^{2} \Omega_\Lambda \sim (2.2{\rm meV})^4$,
where ${\rm meV} = 10^{-3}{\rm eV}$.
Here, $M_{\rm Pl} = (8 \pi G_N)^{-1/2} \sim 2.4\times 10^{30} {\rm meV}$ is 
the (reduced) Planck scale
and $H_0 \sim 1.4\times 10^{-30} {\rm meV}$ is the current 
Hubble parameter. 
$\rho_{\rm{DE}}$ is far smaller than 
the expected magnitude of vacuum energy $\Lambda^4$ in a theory 
with an ultraviolet (UV) cutoff $\Lambda$.
If we take $\Lambda$ to be $M_{\rm Pl}$, $\rho_{\rm{DE}}$ 
is smaller than $\Lambda^4$ by more than 120 orders of magnitude. 
This is the cosmological constant problem~\cite{WeinbergCC}. 

On the other hand, we may try to explain the dark energy
as the Casimir energy in the current universe.
In particular, the vacuum energy 
for fluctuations of massless fields in de Sitter background with Hubble 
parameter $H$ is of order $H^4$, 
and has $w=-1$~\cite{BunchDavies, Dowker}. 
Since our present universe is close to de Sitter space,
one may wonder if dark energy can be explained as 
vacuum energy in de Sitter with the current Hubble 
parameter $H_{0}$, but this does not seem to be plausible. 
Dark energy that we observe as $M_{\rm Pl}^2H_{0}^{2}$ is much larger 
than the expected contribution from a single field $H_{0}^4$. 

However, $H_0$ is not the only dimensionful quantity 
that affects the renormalized energy-momentum tensor (EMT). To compute
the expectation values of fluctuations, we need to specify 
the vacuum state. This may depend on the global properties 
of the geometry and the whole history of the universe,
and thus different scales might be introduced into the problem. 
There is by now strong evidence~\cite{PlanckInf} that there has 
been a period of inflation with the Hubble parameter 
$H_I$ much larger than $H_0$. 
It would be reasonable to take the vacuum to be 
the Bunch-Davies (BD) vacuum~\cite{BunchDavies} 
for de Sitter space with $H_I$.  
Fluctuations of a massless scalar in de Sitter background 
are of order $H_I$. Fluctuations 
are frozen (remain constant) outside the Hubble radius
(see, e.g., Ref.~\cite{Mukhanov:2005sc}); thus, infrared (IR) modes
could have a large value in the universe after inflation. 
In fact, these fluctuations are considered to be the 
origin of the fluctuations in the cosmic microwave background 
(CMB) that are observed today~\cite{PlanckCosm, PlanckInf}. 

In Refs.~\cite{Prokopec1} and \cite{AIS}, the time evolution of the
EMT is calculated for a minimally coupled massless scalar field.
The fluctuations are generated during the inflationary universe and
evolve until the late-time universe of the radiation-dominated (RD) and 
matter-dominated (MD) eras. 
The equation of state $w$ approaches $w=1/3$ and $w=0$
in the RD and MD periods, respectively. 
The magnitude of the present energy density is of order $H_I^{2} H_0^{2}$,
and is still much smaller than that of the dark energy in our universe. 
The analysis is extended to a non-minimally coupled 
scalar field in Ref.~\cite{Prokopec2}. 

In order to circumvent the smallness of energy density, 
we considered a double inflation model in Ref.~\cite{AIS}.
We assumed that there was an inflation 
with a Hubble parameter $H_P$ of the order of the Planck scale $M_{\rm Pl}$ 
(which should be natural in, e.g., Starobinsky inflation~\cite{Starobinsky}) 
before  the usual inflation with $H_I$ started. 
We fix the initial condition of the fields in the
Planckian inflation period by taking a Bunch-Davies vacuum with the Hubble parameter $H_P$, 
and study the time evolution afterwards. 
In this case, the IR mode is enhanced to 
$H_P$, and 
 the present value of vacuum energy becomes of order $H_{P}^{2} H_{0}^{2}$. 
In order to make these large fluctuations consistent with
the observed value of CMB fluctuations, 
the enhancement needs to be restricted in the far-IR modes
whose wavelengths are
 larger than the current Hubble radius $H_0^{-1}$. 
Since the enhanced modes are still out of the horizon now and
the dominant contribution to the EMT is given by
the spacial-derivative parts $|\nabla \phi|^2$,
 the equation of state  is given by
$w=-1/3$ instead of $w=1/3$ or $w=0$ as in ordinary inflation.

In the present paper, we extend our previous analysis in Ref.~\cite{AIS}
 to an (ultra-light) massive scalar field. 
If  the mass is smaller than the Hubble parameter $m \ll H_I$, 
the wave function is greatly enhanced during the inflation, as in the massless case.
The low-momentum modes remain frozen until the mass  becomes larger than the Hubble parameter
in the late universe when the field starts to oscillate. 
Such ultra-light scalars have been studied extensively as  candidates for dark matter
 (see Ref.~\cite{Hill} and papers citing Ref.~\cite{Hill}). 
The scenario has attracted renewed interest since Ref.~\cite{stringaxiverse}. 
For $m \gtrsim 10^{-24} \mbox{eV}$, it is indistinguishable from the standard
cold dark matter \cite{Hlozek}. For lighter mass 
$10^{-32} \mbox{eV} \lesssim m \lesssim 10^{-25.5} \mbox{eV}$, 
its abundance is strongly constrained by the CMB and galaxy-clustering data. 
If its mass is smaller than the current Hubble parameter $H_0 \sim 10^{-33} \mbox{eV}$, 
it could be a candidate for the  
dark energy at present \cite{Kim,KimSemTsu}. If the light particle is an axion-like particle, 
the initial value of the field is set  by the misalignment mechanism,
so the amplitude of the energy density can be chosen by hand.
A similar idea is given by the quintessence scenario 
(for a review, see, e.g., Ref.~\cite{quintessence}), where
 the initial value of the field is also set by hand.
In this paper, we investigate the possibility that the initial field value is dynamically
determined  by the fluctuations generated during the primordial inflation.

A similar proposal was given in Ref.~\cite{Ringeval}.  
Using the saturated  value $3 H_I^4/16 \pi^2$ of the energy density generated by an inflation
with an infinite duration, or an infinite e-folding,
a tiny  Hubble parameter $H_I  \sim {\rm meV}$ is required to explain the observed value 
of the dark energy.
For an inflation with a finite e-folding number $N$, the energy density becomes
 $(m^2 /2) (H_I/2\pi)^2 N$.
For $m \lesssim H_0$ and $H_I \lesssim 10^{-5} M_{\rm Pl}$, which is
required by the CMB observation,
we need a large e-folding number $N$ to make this energy density comparable to
the current dark energy.  
While the lower bound for $N$ was given in  Ref.~\cite{Ringeval} as $N \gtrsim 10^{9}$,
we obtain a slightly different value $N \gtrsim 10^{12}$ 
by taking numerical factors into account.
We also consider a different physical setting where we suppose 
 a period of pre-inflation with a large Hubble parameter
 $H_P \sim M_{\rm Pl}$ before the ordinary inflation starts.
 In this case, 
an e-folding number $N \sim 240$ for the Planckian inflation is sufficient. 

The purpose of the present paper is twofold.
One is, as we mentioned above, to give conditions to explain the present dark energy by 
the vacuum fluctuations generated during the inflation or pre-inflation,
and to investigate the time evolution of the EMT 
through the RD and MD eras.
We extend our analyses to see the behavior in the future
when $m>H$ and the EMT behaves as a dust with an oscillating $w$,
though the back reaction of the induced EMT to the geometry
needs to be included.
The second purpose is to
obtain the {\it exact} wave function of a massive scalar field with the Bunch-Davies initial condition,
for all the time and for all the momenta,
and  to calculate the EMT by using it.
While analyses using only zero-momentum modes are often performed in the literature, 
our analyses contain nonzero-momentum modes as well.
We also obtain some approximated forms of the wave function
by applying the WKB approximation,
and by using the power-expansion and asymptotic forms of the special functions
describing the exact wave function.
These results will serve as bases for future calculations,
e.g., when interactions among different modes become important.
Studies for the former purpose are mainly given in section~\ref{sec:dark_energy},
and those for the latter in sections~ \ref{sec:scalar} and \ref{sec:EMT}.

Our results will shed light on the effect of 
almost massless and non-interacting scalar fields, such as axions.
Although our work is related to the studies of fluctuations of 
gravitons or inflatons \cite{Abramo1, Abramo2, Abramo3, Kolb, Barausse},  
further modifications are necessary since
the EMT for gravitons has different
tensor structures from the scalar fields.
One should also study quantum fluctuations around the classical value of the inflaton field
developed in the potential.
We believe that our work serves as a starting point for 
a study of the effects of those fields.

The paper is organized as follows.
In section~\ref{sec:scalar}, we solve the equation of motion of a massive scalar field
in the history of the universe with the inflation period followed by
the RD universe.  
In section~\ref{sec:EMT}, we calculate the EMT and study its time evolution.
In section~\ref{sec:dark_energy}, we  obtain the conditions in which the vacuum fluctuation of the ultra-light scalar
 explains the dark energy at present. 
The last section, section~\ref{sec:concl}, is devoted to conclusions and discussions.
In Appendix~\ref{sec:IRmodes}, we calculate the time evolution of the EMT by using the zero-momentum
approximation. The time evolution in the MD period is given by using this approximation.

%%%%%%%%%%%%%%%%%%%%%%%%%%%%%%%%
\section{Massive scalars in an expanding universe}
\label{sec:scalar}
\setcounter{equation}{0}

Our universe is approximated by the Robertson-Walker spacetime
with the inflation, RD (radiation-dominated), and MD (matter-dominated) periods.
In this section, we focus on the first two stages of the universe
and study the detailed behaviors of the wave function of a  massive scalar field.\footnote{In
\cite{Padmanabhan}, particle content and the degree of classicality were studied in a similar background
(de Sitter in inflation, followed by RD, and late-time de Sitter).} 
The metric is given by
$ds^2 = a(\eta)^2\left[d\eta^2 -(dx^i)^2\right] $ 
in terms of the conformal coordinates $(\eta, x^i)$, and 
the scale factor $a(\eta)$ is given by
\beq
a(\eta)=\left\{ \begin{array}{lll}
a_{\rm Inf}(\eta)=-\frac{1}{H_I \eta} &(\eta_{\rm ini}<\eta<\eta_1<0) &(\mbox{Inflation}) \\
a_{\rm RD}(\eta)=\alpha \eta &(0<\eta_2<\eta) &(\mbox{RD})  
\end{array} \right. \ ,
\eeq  
where $\eta_{\rm ini}$ and $\eta_1$ denote the beginning and end of inflation,
and $\eta_2$ the beginning of the RD period.  
The continuity conditions for the scale factor $a$ and its derivative $a^\prime = \partial_\eta a$
at the boundary of the inflation and RD periods give the relations
\beq
\eta_2 = -\eta_1 \ , \ \  \alpha= \frac{1}{H_I \eta_1^2} \ .
\eeq
The Hubble parameter $H=a'/a^2$ is given by
\beq
H(\eta)=\left\{ \begin{array}{ll}
H_{\rm Inf}(\eta)=H_I  &(\mbox{Inflation})  \\
H_{\rm RD}(\eta)=\frac{1}{\alpha \eta^2}=H_I \left(\frac{\eta_2}{\eta} \right)^2 &(\mbox{RD})  
\end{array} \right. \ .
\eeq  
The CMB fluctuations give a constraint  $H_I<3.6 \times 10^{-5} M_{\rm Pl}.$ 

We consider a minimally coupled massive scalar field $\phi$  with a mass $m$.
Quantum fields are expanded as
\beq
\phi(\eta,x^i) = \int \frac{d^{3}k}{(2 \pi)^{3}}
\left[a_{\bf k} u_{\bf k}(\eta) + a_{-{\bf k}}^\dagger u_{-{\bf k}}(\eta)^*\right]e^{i{\bf k}\cdot {\bf x}} \ ,
\label{phiexpau}
\eeq
where the mode functions
$u_{\bf k}(\eta)$ with the comoving momentum ${\bf k}$ 
are the solutions of the equation of motion, 
$
(\Box+m^2 ) u =0 ,
$
and are
chosen to asymptote to positive-frequency modes in the remote past.
A vacuum is then defined by $a_{\bf k} |0\rangle =0$.
The vacuum $|0 \rangle$, which is an in-state, evolves as $\eta$ increases, and
if an adiabatic condition is  broken, the state gets excited above an adiabatic ground
state at each moment $\eta$.

In the Robertson-Walker spacetime,
the wave equation for $\chi_{\bf k}(\eta) \equiv a(\eta) u_{\bf k}(\eta)$ is given by
\beq
\left[ -\partial_\eta^2 + \frac{1}{6} R a^2 - m^2 a^2 \right] \chi_{\bf k}(\eta)
= k^2 \chi_{\bf k}(\eta) \ ,
\label{waveeq}
\eeq
where $k=\sqrt{{\bf k^2}}$ and 
\beq
\frac{1}{6}Ra^2 = \frac{a''}{a}
=\left\{\begin{array}{ll}
2/\eta^2   & \mbox{(Inflation)} \\
0   & \mbox{(RD)} \\
\end{array}
\right. \ .
\label{potIRM}
\eeq

Since the wave equation (\ref{waveeq}) 
has a form of the Schr\"{o}dinger equation 
$[-\partial_\eta^2 + V(\eta) ]\chi(\eta) =E \chi(\eta)$, 
the  Klein-Gordon (KG) inner product
\beq
(\chi_1, \chi_2)_\eta=  i \left(\chi_1^*  (\partial_\eta \chi_2) - (\partial_\eta \chi_1^*) \chi_2\right) 
\eeq
is preserved. We normalize the wave functions $\chi(\eta)$
in terms of the KG inner product.

%%%%%%%%%%%%%%%%%%%%%%%%
\subsection{Wave functions in the inflationary period}
In the inflationary period, 
a solution of the wave equation is given by
\beq
\chi_{\rm BD} (\eta)= \frac{\sqrt{-\pi \eta}}{2} e^{i (2 \nu+1) \pi/ 4} 
H_\nu^{(1)} (-k \eta) \ ,
\label{BDwavefunc}
\eeq
where $H_\nu^{(1)}$ is the Hankel function of the first kind and
\beq
\nu=\sqrt{\frac{9}{4} -\left( \frac{m}{H_I} \right)^2} \simeq \frac{3}{2}-\frac{1}{3} \left( \frac{m}{H_I} \right)^2 >0
\eeq
for $m \ll H_I.$
Another solution is given by its complex conjugate.
Using an asymptotic form of the Hankel function at $|z| \rightarrow \infty$,
\beq
H_\nu^{(1)}(z) \rightarrow \sqrt{\frac{2}{\pi z}} e^{i (z-(2 \nu+1) \pi/4)} \ ,
 \eeq
$\chi_{\rm BD}(\eta)$ is shown to have the Bunch-Davies initial condition 
\beq
\chi_{\rm BD}(\eta) \rightarrow \frac{e^{- ik \eta}}{\sqrt{2k}}
\eeq 
 at $\eta \rightarrow - \infty$.

For $\eta \rightarrow 0$, on the other hand, the wave function can be approximated 
by using the expansion formula of the Hankel function 
$H_\nu^{(1)}(z)=J_\nu(z) + i Y_\nu(z)$ 
near $z =0$,
\beq
J_\nu(z) \simeq \frac{1}{\Gamma(\nu+1)} \left( \frac{z}{2} \right)^{\nu} \ , \ \ 
Y_\nu(z) \simeq \frac{-1}{\sin (\pi \nu) \Gamma(-\nu+1) } \left( \frac{z}{2} \right)^{-\nu} \ , 
\label{bessel_z=0}
\eeq
as
\beq
\chi_{\rm BD}(\eta) \simeq 
 \frac{i }{\sqrt{2}} k^{-\frac{3}{2}+ \frac{1}{3} \left( \frac{m}{H_I} \right)^2} (- \eta)^{-1+\frac{1}{3} \left( \frac{m}{H_I} \right)^2}  \ .
 \label{uBDIRappr}
\eeq
This is valid when $k |\eta| = k_{\rm phy} /H_I <1$ is satisfied,
where the physical momentum in the inflationary period is given by
$k_{\rm phy} =k/a=k H_I |\eta|$.
In the massless limit, the wave function 
$u_{\rm BD}(\eta) = \chi_{\rm BD}(\eta)/a_{\rm Inf}(\eta) $
becomes almost constant in time $\eta$, and behaves as $k^{-3/2}$.
This is consistent with the $\eta \rightarrow 0$ behavior of  the
exact massless wave function:
\beq
\chi_{\rm BD,m=0} =\frac{1}{\sqrt{2k}} \left( 1 - \frac{i}{k \eta} \right) e^{-i k \eta} \ .
\eeq

%%%%%%%%%%%%%%%%%%%%%%%%%%%
\subsection{Wave functions in the RD period}

In the RD period, the wave equation (\ref{waveeq}) is written as
 \beq
\left[ \partial_x^2 + (q^2 + \frac{x^2}{4}) \right] \chi = 0 \ ,
\label{waveeqRDxq}
 \eeq
where we have rescaled the parameters $\eta$ and $k$ as 
\beq
x \equiv (2 \tilde{m})^{1/2} \eta \ , \ \ 
q \equiv (2 \tilde{m})^{-1/2} k \ , \ \ 
\tilde{m} \equiv \frac{m}{H_I \eta_1^2} \ .
\eeq
Since the Hubble parameter is given by $H=H_I (\eta_1/\eta)^2$, 
the new parameters $x$ and $q$ have the
following physical meaning:
\beq
x=\sqrt{\frac{2m}{H}} \ , \ \ qx =k \eta= \frac{k_{\rm phy}}{H} \ ,  \ \
\frac{q}{x}=\frac{k_{\rm phy}}{2m} \ ,
\label{xqphys}
\eeq
where $k_{\rm phy}=k/a_{\rm RD}$ is the physical momentum in the RD period.
Equation (\ref{waveeqRDxq}) is nothing but 
the Schr\"{o}dinger equation in an inverted harmonic oscillator 
with the Hamiltonian $H=-\partial_x^2-x^2/4$ and the energy eigenvalue $q^2$. 
 
The exact solution can be obtained by analytical continuation of the solution in the harmonic potential, 
and is given by
 \beq
 \chi_{\rm EX} (x) = c e^{-\pi q^2/4} D_{-1/2-i q^2}(e^{i \pi/4}x) \ ,
 \label{RDexactsol}
 \eeq
where $c=(2 \tilde{m})^{-1/4}$.
$D_{n}(z)$ is the parabolic cylinder function 
satisfying the Weber differential equation 
\beq
\left[\partial_z^2 + (n+1/2-z^2/4) \right] D_{n}(z) =0 \ .
\eeq
It has the following asymptotic behavior
 \beq
 D_n(z) \rightarrow e^{-z^2/4} z^n \ \ \ \mbox{for} \ |z| \rightarrow \infty \ .
 \label{asympPC}
 \eeq
The exact solution (\ref{RDexactsol}) approaches the WKB-approximated one (\ref{WKBqltx})
for $x \rightarrow \infty$, as will be seen below.
 
 \vspace{5mm}
 
 If the adiabaticity condition is satisfied,  the exact wave function
 is approximated by the WKB wave function:
 \beq
 \chi_{\rm WKB} (x) =c  \frac{e^{-i \int \omega dx}}{\sqrt{2 \omega}} \ ,  \ \  
\omega= \sqrt{\frac{x^2}{4} + q^2} \ .
\label{WKBfunc}
 \eeq
 The integral of $\omega(x)$ is given by
 \beq
 \int^x \omega (x) dx = \frac{x}{4} \sqrt{x^2+ 4q^2} + q^2 \ln (x+\sqrt{x^2+4 q^2}) 
  -q^2 (\ln 2 +\frac{1}{2}) + \frac{\pi}{8} \ ,
  \label{integralomega}
 \eeq
where the  $x$-independent terms are fixed by comparing them with the 
asymptotic behavior of the exact wave function.
By taking $c=(2 \tilde{m})^{-1/4}$, the WKB wave function (\ref{WKBfunc}) is normalized 
by the KG inner product with respect to the $\eta$-derivative as
$( \chi_{\rm WKB} , \chi_{\rm WKB} )_\eta =1$.

\begin{figure}[t]
 \begin{center}
   \includegraphics[width=7cm]{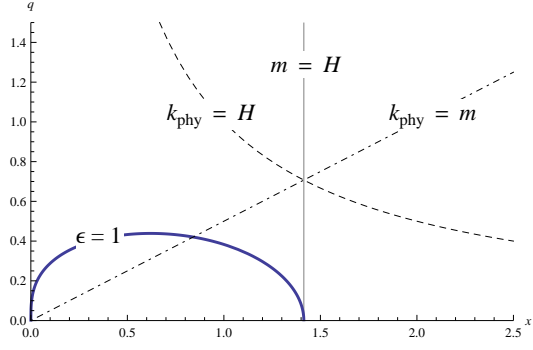}
   \caption{The wave function in the RD period changes its behavior
 depending on the parameters $x=\sqrt{2m/H}$ (time) and $q=k_{\rm phy}/\sqrt{2mH}$ (momentum).
The thick solid curve represents $\epsilon = 1$;
   outside the curve, the WKB approximation is valid.
   The vertical line is $x=\sqrt{2}$, which corresponds to $m=H$ due to (\ref{xqphys});
   to the left of the line, $m<H$.
   The dashed curve is $qx=k_{\rm phy}/H=1$;
   under the curve, $k_{\rm phy}<H$.
   The dot-dashed line is $q=x/2$, which corresponds to $k_{\rm phy}=m$;
    under the line, $k_{\rm phy}<m$.}
   \label{fig-WKB}
 \end{center}
\end{figure}

The adiabaticity condition is given by
 \beq
\epsilon \equiv \left| \frac{\omega'(x)}{\omega^2} \right|= 
\frac{x }{4 \left( q^2 + x^2/4 \right)^{3/2}}
 < 1 \  .
 \eeq
In Figure~\ref{fig-WKB}, the curve $\epsilon=1$ is drawn on the $(x,q)$ plane by a thick solid line.
Outside  the semicircle, the WKB approximation is valid. 
Also shown in the figure are 
the vertical line $m=H$ at $x=\sqrt{2}$, the curve $k_{\rm phy}=H$ (dashed line), 
and the tilted line $k_{\rm phy}=m$ (dot-dashed line).
%%%%%%%%%%%%%%%%%%%%%%%%%%%%%
%%%%%%%%%%%%%%%%%%%%%%%%%%%%%%%%

For $q<x/2$, namely, $k_{\rm phy}<m$, the WKB wave function (\ref{WKBfunc})
is further approximated as
 \beq
 \chi_{\rm WKB} \simeq c \frac{e^{-i x^2/4-i q^2 \ln x-i \pi/8} }{\sqrt{x}} \ .
 \label{WKBqltx}
 \eeq
Then the exact wave function (\ref{RDexactsol}) 
can be shown to  asymptote to the WKB wave function by using (\ref{asympPC}). 
Note also that $x^2/4= m/(2H)= mt$ where $t$ is the physical time in the RD period,
and the wave function oscillates as $e^{-imt}$.
 
For $q>x/2$ (i.e., $k_{\rm phy}> m$), 
the WKB wave function is reduced to the plane wave  
\beq
\chi_{\rm WKB} \simeq c e^{-i(q^2 \ln q -q^2/2 +\pi/8)}\frac{e^{-i qx}}{\sqrt{2q}} \ .
\label{WKBqgtx}
\eeq  
Since $qx=k_{\rm phy}/H=2k_{\rm phy} t $,
the oscillation behavior $ e^{-2 i k_{\rm phy} t}$ is controlled by the momentum, not by the mass.

\vspace{5mm}

For small $x$,
we can approximate the exact wave function $\chi_{\rm EX}$ of
 (\ref{RDexactsol})  by a power series of  $x$.
The parabolic cylinder function is expanded as 
\beq
D_{-1/2- i q^2}(e^{i \pi/4} x) = \sum_{r=0}^\infty  c_{r} (q^2)  x^{r} \ ,
\eeq
where the first two coefficient functions are given by
\beq
c_0(q^2) = \frac{2^{-1/4-i q^2/2} \sqrt{\pi}}{\Gamma(3/4+iq^2/2)}  \ , \ \ 
c_1(q^2) =- e^{i \pi/4} \frac{2^{1/4-i q^2/2} \sqrt{\pi}}{\Gamma(1/4+iq^2/2)}  \ .
\label{c0c1}
\eeq
The rest of the coefficient functions $c_r (q^2)$ are related 
to  them as, e.g.,
\beqa
&& c_2(q^2) =-\frac{q^2}{2!} c_0(q^2) \ , \ \  c_3(q^2)= -\frac{q^2}{3!}  c_1(q^2) \ , \n 
&&c_4(q^2)=\frac{-1/2+ q^4}{4!} c_0(q^2) \ , \ \ c_5(q^2) =\frac{-3/2 +  q^4}{ 5!}  c_1(q^2) \ , \n
&&c_6(q^2)=q^2 (\frac{7/2 - q^4}{6!})c_0(q^2) \ , \ \ 
c_7(q^2)= q^2 (\frac{13/2 - q^4}{7!} ) c_1(q^2) \ .
\eeqa
If we take the highest powers of $q^2$ in each $c_r(q^2)/c_0(q^2)$ and $c_r(q^2)/c_1(q^2)$, 
the parabolic cylinder function is expanded as
\beqa
&& D_{-1/2- i q^2}(e^{i \pi/4} x) 
= c_0(q^2)\Bigg[\cos (qx) +\left(-\frac{1}{2\cdot 4!} x^4+{\cal O}(x^8) \right) 
+\left(\frac{7}{2\cdot 6!} x^4+{\cal O}(x^8) \right) (qx)^2
\n
&&+ \left( -\frac{11}{8!}x^4 + {\cal O}(x^8) \right) (qx)^4
+ \left( \frac{25}{10!}x^4 + {\cal O}(x^8) \right) (qx)^6 +\cdots
 \Bigg]  \n
&&+ c_1(q^2) x\Bigg[\frac{\sin (qx)}{qx} +\left(- \frac{3}{2\cdot 5!} x^4+{\cal O}(x^8) \right) 
+\left(\frac{13}{2\cdot 7!}x^4 +{\cal O}(x^8) \right) (qx)^2
\n
&&+ \left( -\frac{17}{9!}x^4 + {\cal O}(x^8) \right) (qx)^4
+ \left( \frac{35}{11!}x^4 + {\cal O}(x^8) \right) (qx)^6 + \cdots
\Bigg] \ . 
\label{Dexp} 
\eeqa
For small $x$, it can be approximated as
\beq
D_{-1/2- i q^2}(e^{i \pi/4} x)  \simeq c_0(q^2) \cos(qx) + c_1(q^2) \frac{\sin(qx)}{q}  \  .
\label{Dapproxsmall}
\eeq
Since the above expansion turns out to be a double expansion of 
$x^4$ and $(qx)^2$, it is  a good approximation for small $q^2$ at $x<1.$
At the same time, since we took the highest orders of $q^2$ in
each  $c_r(q^2)$, it should also be a good approximation for large $q^2$. 
Indeed, we have confirmed numerically that the remaining terms in the square brackets in (\ref{Dexp})
decrease and oscillate as a function of $qx$.
This will be related to the fact that the expansion of  $(qx)^2$ is
an alternative power series. Hence, (\ref{Dapproxsmall})  
gives a good approximation 
for small $x\lesssim 1$, irrespective of the magnitude of $q$.

%%%%%%%%%%%%%%%%%%%%%%%%%%%%%%%
\subsection{Determination of the Bogoliubov coefficients} 
We now solve the wave equation throughout the inflationary and RD periods by
imposing the BD initial condition.
The wave function can be written as
\beq
\chi(\eta)=\left\{ \begin{array}{ll}
\chi_{{\rm BD}}(\eta)  &(\mbox{Inflation})  \\
A(k) \chi_{{\rm EX}}(\eta) + B(k) \chi_{{\rm EX}}^*(\eta) &(\mbox{RD})  
\end{array} \right. \ ,
\label{chiInfRDlc}
\eeq  
where $\chi_{\rm BD}(\eta)$ and $\chi_{\rm EX}(\eta)$
are defined in (\ref{BDwavefunc}) and (\ref{RDexactsol}).\footnote{
It is an abuse of notation, but we write the wave function 
in the RD period as $\chi_{\rm EX}(\eta)$  instead of $\chi_{\rm EX}(\sqrt{2\tilde{m}}\eta)$
for notational simplicity.} 
The Bogoliubov coefficients $A(k)$ and  $B(k)$ can be determined by
imposing  continuity of $\chi$ and $\partial_\eta \chi$ at the boundary 
of the inflationary and RD periods, and are given by
\beqa
A(k) &=& i \left[ \chi_{\rm EX}^* (\eta_2) \cdot \partial_\eta  \chi_{\rm BD}  (\eta_1) 
- \partial_\eta \chi_{\rm EX}^*  (\eta_2)  \cdot \chi_{\rm BD} (\eta_1) \right] \ , \n
B(k) &=& -i \left[ \chi_{\rm EX} (\eta_2) \cdot  \partial_\eta   \chi_{\rm BD} (\eta_1) 
- \partial_\eta \chi_{\rm EX}  (\eta_2) \cdot \chi_{\rm BD} (\eta_1) \right]  \ ,
\label{ABchiEXchiBD}
\eeqa
where the KG normalization of the wave function
$i[\chi_{\rm EX}^* (\eta_2) \cdot \partial_\eta  \chi_{\rm EX}  (\eta_2) 
- \partial_\eta \chi_{\rm EX}^*  (\eta_2)  \cdot \chi_{\rm EX} (\eta_2)] =1$ is used.
These coefficients satisfy the relation $|A(k)|^2-|B(k)|^2=1$.

The Bogoliubov coefficients $A(k)$ and $B(k)$
in the IR region $k|\eta_1| <1$
are calculated by
using the wave function $\chi_{\rm BD}(\eta)$ 
in (\ref{uBDIRappr}). 
Its derivative is written as
\beq
\partial_\eta \chi_{\rm BD}(\eta) \simeq \frac{1-\frac{1}{3} \left( \frac{m}{H_I} \right)^2}{-\eta} \chi_{\rm BD}(\eta)
\simeq \frac{\chi_{\rm BD}(\eta)}{-\eta} \ ,
\eeq
where $(m/H_I) \ll 1$ was used in the second equality,
and (\ref{ABchiEXchiBD}) become
\beqa
A(k) 
&\simeq& 
\left( 
 \frac{\chi_{\rm EX}^* (\eta_2)}{-\eta_1}
- \partial_\eta \chi_{\rm EX}^* (\eta_2)
\right) (i \chi_{\rm BD} (\eta_1) )  \ , \n
B(k) &\simeq& -A(k)^* \ .
\label{ABapprBD}
\eeqa
Note that $i \chi_{\rm BD}(\eta_1)$ is real within this approximation.

Since, at $\eta=\eta_2$, $x= \sqrt{2 \tilde{m} } \eta_2=\sqrt{2m/H_I} \ll 1$
is satisfied, we can use the approximation (\ref{Dapproxsmall}) for 
 $\chi_{\rm EX}(\eta_2)$.
Then the coefficient $A(k)$ in (\ref{ABapprBD}) becomes
\beqa
A(k)
&\simeq & \frac{ i e^{- \frac{\pi k^2}{8 \tilde{m}}}   \chi_{\rm BD} (\eta_1)}{ (2 \tilde{m})^{1/4} }
\frac{c_0^*}{|\eta_1|}
\left[    (\cos(k \eta_2) + k\eta_2 \sin(k \eta_2) ) 
+ \sqrt{2\tilde{m}} \frac{c_1^*}{c_0^*}  (\frac{\sin(k \eta_2)}{k } - \eta_2 \cos(k \eta_2)) 
\right] \n
&\simeq&
\frac{ i e^{- \frac{\pi k^2}{8 \tilde{m}}}   \chi_{\rm BD} (\eta_1)}{ (2 \tilde{m})^{1/4} }
\frac{c^*_0}{|\eta_1|}  \ ,  
\label{BogCapprox}
\eeqa
where $k\eta_2 < 1$ and $\sqrt{2\tilde{m}} \eta_2 \ll 1$ were used in the second equality.

%%%%%%%%%%%%%%%%%%%%%%%%%%%%%%%%
\subsection{Behaviors of the wave functions with BD initial condition}
By using the  Bogoliubov coefficients $A(k)$ and $B(k)$, the wave function $u(\eta)$ 
in the RD period with the BD initial condition is given by
\beqa
u(\eta)&=& -\frac{i H_I}{\sqrt{2} }
k^{- \frac{3}{2}+\frac{1}{3} \left(\frac{m}{H_I} \right)^2} 
(-\eta_1)^{\frac{1}{3} \left(\frac{m}{H_I} \right)^2} \n
&\times&  
\frac{e^{-\pi k^2/4 \tilde{m}}}{\sqrt{2 \tilde{m} } \eta}
2 \Im \Bigl[
c^*_0 \   D_{-\frac{1}{2}-i \frac{k^2}{2\tilde{m}}}(e^{i \pi/4} \sqrt{2 \tilde{m}} \eta) \Bigr] \ ,
\label{uwavefunction}
\eeqa
where $c_0$ is defined in (\ref{c0c1}).
This is one of the main results of the paper, and gives a starting point to calculate the
EMT in the RD period.
This expression is valid for IR modes with  momentum $k|\eta_1|=k\eta_2 < 1$.
For UV modes with $k|\eta_1|>1$, the wave function is not enhanced like this. 

For early times $x=\sqrt{2\tilde{m}}\eta=\sqrt{2m/H}\lesssim1$, 
the exact solution $\chi_{\rm EX}(\eta)$ is approximated by (\ref{Dapproxsmall}),
and the  wave function $u(\eta)$ is simplified as
\beq
u(\eta) \simeq  \frac{i H_I}{ \sqrt{2} }
k^{- \frac{3}{2}+\frac{1}{3} \left(\frac{m}{H_I} \right)^2} 
(-\eta_1)^{\frac{1}{3} \left(\frac{m}{H_I} \right)^2} \ 
\ \frac{\sin(k \eta)}{k\eta}  \   .
\label{uwaveBB}
\eeq
Here we have used the identity
\beq
2 e^{-\pi k^2/4 \tilde{m}}    \Im (c^*_0 c_1) =-1 \ ,
\eeq
which can be proved by using
the property of the KG inner product $(\chi_{\rm EX}, \chi_{\rm EX})_\eta=1$
at $\eta=0.$
For $k \eta < 1$,  $u(\eta)$ becomes almost constant, i.e., {\it frozen}.

For later times $x\gtrsim \sqrt{2}$, the WKB approximation becomes
valid, and the wave function 
$u(\eta)$ is written as
\beqa
u(x) &\simeq&  -\frac{i H_I(2\tilde{m})^{- \frac{3}{4}+\frac{1}{6} \left(\frac{m}{H_I} \right)^2} }{ \sqrt{2} }
q^{- \frac{3}{2}+\frac{1}{3} \left(\frac{m}{H_I} \right)^2} 
(-\eta_1)^{\frac{1}{3} \left(\frac{m}{H_I} \right)^2} \n
&& \times~\frac{e^{-\pi q^2/4} }{x }
2 \Im \left[c_0^* \frac{e^{-i \int^x \omega dx}  }{ \sqrt{2 \omega} }
\right] \ .
\label{uwaveWKB}
\eeqa
The WKB approximation is also valid for the UV modes 
(outside of the semicircle in Figure~\ref{fig-WKB}) even at early times $x<\sqrt{2}$.

As we saw previously,
the WKB wave function has two different behaviors, 
(\ref{WKBqgtx})  at $q>x/2$ ($k_{\rm phy} > m$), and  (\ref{WKBqltx}) at
$q<x/2$ ($k_{\rm phy} < m$). 
In the region $q>x/2$, 
it can be shown  numerically that 
the $x$-independent phase in the square brackets in (\ref{uwaveWKB}) 
vanishes for large $q >1$
and  (\ref{uwaveWKB}) becomes identical with (\ref{uwaveBB}).
In contrast, for $q<x/2$, the WKB wave function
 (\ref{WKBqltx})
oscillates as $\chi_{\rm WKB} \sim e^{-i x^2/4}=e^{-imt}$.

\begin{figure}[t]
 \begin{center}
   \includegraphics[width=7cm]{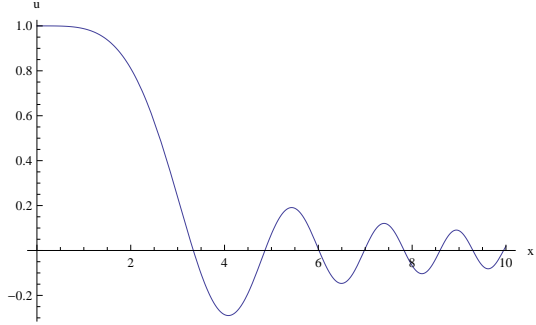}
   \caption{
The behavior of the wave function $u(x)$ in
(\ref{uwavefunction}), normalized by
 $(i H_I/\sqrt{2} )
k^{- \frac{3}{2}+\frac{1}{3} \left(\frac{m}{H_I} \right)^2} 
(-\eta_1)^{\frac{1}{3} \left(\frac{m}{H_I} \right)^2}$.
The momentum is set to $k = 0$.
The wave function is frozen until $x\sim \sqrt{2}$ (i.e., $m\sim H$) and then starts to oscillate. 
}
   \label{fig:uatp0}
 \end{center}
\end{figure}

In Figure~\ref{fig:uatp0}, the behavior of the wave function of 
the exact solution in (\ref{uwavefunction}) is plotted for $q=0$. 
As expected,
it is almost constant (frozen) for $x \lesssim \sqrt{2}$, and 
slowly decreases with an oscillation  for $x \gtrsim \sqrt{2}$.

%%%%%%%%%%%%%%%%%%%%%%%%%%%%%%%%%
\section{Evolution of the energy-momentum tensor}
\label{sec:EMT}
\setcounter{equation}{0}
In this section, we calculate the EMT (energy-momentum tensor) in the RD period.
The physical results that are relevant to the application to the dark energy 
are summarized in  section~\ref{sec:dark_energy}. 

The vacuum expectation value of the EMT  is given by
\beqa
\rho(\eta)
&=& \frac{1}{a^2} \int \frac{d^{3}k}{(2 \pi)^{3}} \frac{1}{2}
\left[ |u'_k(\eta)|^2 + k^2 |u_k(\eta)|^2  + (ma)^2 |u_k(\eta)|^2 \right] \ , 
\label{rhogenu} \\
p(\eta)
&=& \frac{1}{a^2} \int \frac{d^{3}k}{(2 \pi)^{3}} \frac{1}{2}
\left[ |u'_k(\eta)|^2 -\frac{1}{3} k^2 |u_k(\eta)|^2 - (ma)^2 |u_k(\eta)|^2 \right] \  .
\label{pgenu}
\eeqa
The first term in $\rho$ and $p$ is the time-derivative term, which gives the equation of state $w=p/\rho=1$.
The second term is the spacial-derivative term, and gives $w=-1/3$.
The third term is a contribution of the mass term, and gives $w=-1$.

The integrals (\ref{rhogenu}) and (\ref{pgenu}) are divergent in the UV region,
and need subtraction of the UV divergences.
After the subtraction is performed,
we  simply cut off the integral at the momentum $k=1/|\eta_1|$,
since only those IR modes are enhanced during the inflationary period.
(For more details, see Section~6 of Ref.~\cite{AIS}.)
The conformal anomaly is given by the subtraction term,
but its contribution to the EMT is of the order of $H^4$
and is negligibly small compared to the vacuum fluctuations generated during the inflation
of the order of $H_I^2 H^2$, so we will not consider it in the present paper.

%%%%%%%%%%%%%%%%%%%%
\subsection{Massless case}
Before investigating the massive scalar field, 
let us first summarize the evolution of the EMT in the massless case
studied in Refs.~\cite{Prokopec1,AIS}. 
The wave function with the BD initial condition is given by
\beq
u(\eta) \simeq \frac{i H_I}{\sqrt{2}} k^{-3/2} \frac{\sin (k \eta)}{k \eta} \ ,
\eeq
for $k |\eta_1| <1$.
Then the energy and pressure  densities become
\beqa
\rho(\eta) &\simeq& \frac{H_I^2}{8\pi^2 a^2} \int^{1/|\eta_1|}_{1/|\eta_{\rm ini}|} k dk 
\left[ (\partial_y f(y))^2+f(y)^2 \right]_{y=k\eta} \ , \n
p(\eta )&\simeq& \frac{H_I^2}{8\pi^2 a^2} \int^{1/|\eta_1|}_{1/|\eta_{\rm ini}|} k dk 
\left[ (\partial_y f(y))^2- \frac{f(y)^2}{3} \right]_{y=k\eta} \ ,
\label{EMTmassless}
\eeqa
with $f(y)=\sin (y)/y.$
 The $k$-integral is performed for  momenta with which the wave function
is enhanced during the inflation. Hence, if the inflation continues during
$\eta \in [\eta_{\rm ini}, \eta_1]$, the
integral region of the comoving momentum
is restricted in $k \in [1/|\eta_{\rm ini}|, 1/|\eta_1|]$. 
There is no IR divergence in (\ref{EMTmassless}), and the IR cutoff does not play an important role.

In the RD period, some of the enhanced modes enter the horizon again.
The modes with $k < 1/\eta$ are still out of the horizon and are frozen.
Then the time-derivative term, the first term in (\ref{EMTmassless}), vanishes and 
the spacial-derivative term, the second term in (\ref{EMTmassless}), gives
\beq
\rho^{\rm IR}(\eta) \simeq \frac{H_I^2}{8 \pi^2 a^2} \int^{1/\eta} k dk
= \frac{H_I^2}{16\pi^2 a^2 \eta^2} =\frac{H_I^2 H^2}{16\pi^2} \ .
\label{EMTmasslessIR}
\eeq 
On the other hand, 
the UV modes with $k > 1/\eta$ have already entered the horizon
and the wave functions are time dependent.
In performing the $k$-integration over $[1/\eta, 1/\eta_2]$,
we estimate the oscillating integrals as
\beq
 \int  \frac{dy}{y} \sin^2(y)  \ , \ \int \frac{dy}{y} \cos^2(y)
 \approx \int  \frac{dy}{2y} \ .
\eeq  
Then the energy and pressure densities 
contributed from the UV modes become
\beqa
\rho^{\rm UV}(\eta) 
&\simeq & \frac{H_I^2}{8 \pi^2 a^2} \int_{1/\eta}^{1/\eta_2} \frac{k dk}{(k\eta)^2} 
=  \frac{H_I^2 H^2}{8 \pi^2} \ln \left(\frac{\eta}{\eta_2} \right) 
=  \frac{H_I^2 H^2}{8 \pi^2} N_{\rm RD} \ , \n
p^{\rm UV} (\eta)&\simeq& 
\frac{H_I^2}{8 \pi^2 a^2} \int_{1/\eta}^{1/\eta_2} \frac{k dk}{(k\eta)^2} \left(\cos^2 (k\eta) -\frac{\sin^2 (k \eta)}{3} \right)
\simeq \frac{\rho^{\rm UV}}{3} \ .
\label{EMTmasslessUV}
\eeqa
Here we have dropped higher-order terms with respect to $(k \eta)^{-1}$.
We have also defined an e-folding during the RD period,
\beq
N_{\rm RD}  = \ln \frac{\eta}{\eta_2} =\ln \frac{a(\eta)}{a_{\rm BB}} \ ,
\label{N_RD}
\eeq
where BB stands for  the big bang; namely, 
$a_{\rm BB}$ means the scale factor at the beginning of the RD period
or the end of inflation. 
$N_{\rm RD}$ represents the number of degrees of freedom
 that were enhanced during
the inflation and have already entered the horizon.

The total energy density is given by $\rho=\rho^{\rm IR}+\rho^{\rm UV}$, but
 the UV contribution (\ref{EMTmasslessUV}) 
dominates the IR part (\ref{EMTmasslessIR}) when $N_{\rm RD} \gtrsim 1$.
Thus the equation of state of the vacuum fluctuations
generated during the inflation approaches $w=1/3$ 
soon after the RD period begins.

%%%%%%%%%%%%%%%%%%%%%%%%%%%
\subsection{Massive case at early times}
\label{sec:massive_early}
We now study the evolution of the EMT in the massive case.
In terms of the variables $x$ and $q$, (\ref{rhogenu}) and (\ref{pgenu}) become 
\beqa
\rho(\eta)
&=& \frac{\sqrt{2}\tilde{m}^{5/2}}{\pi^2 a^2} \int^{1/x_1}_{1/x_{\rm ini}} q^2 dq
\left[ |\partial_x u|^2 + (q^2 +\frac{x^2}{4} )|u|^2   \right] \  , 
\label{rhogenup} \\
p(\eta)
&=& \frac{\sqrt{2}\tilde{m}^{5/2}}{\pi^2 a^2} \int^{1/x_1}_{1/x_{\rm ini}} q^2 dq
\left[  |\partial_x u|^2 - (\frac{q^2}{3} +\frac{x^2}{4} )|u|^2
 \right] \ ,
\label{pgenup}
\eeqa
where  $x_{\rm ini}=\sqrt{2 \tilde{m}} |\eta_{\rm ini}|$
and $x_1=\sqrt{2 \tilde{m}} |\eta_1|=\sqrt{2m/H_I}$ correspond to the beginning and end of the inflation period.
The integration region represents the momentum region 
where the wave function is amplified.

We first consider the behavior of the EMT at early times 
$x=\sqrt{2m/H} <1$. 
The wave function  is  approximated by (\ref{uwaveBB}), and the EMT becomes
\beqa
\rho &\simeq& \frac{H_I^2 \tilde{m} x_1^{\frac{2}{3} \left(\frac{m}{H_I} \right)^2}}{4 \pi^2 a^2 } \int^{1/x_1}_{1/x_{\rm ini}} dq \
q^{1+\frac{2}{3} \left(\frac{m}{H_I} \right)^2 } 
\left[(\partial_y f(y))^2 + (1+\frac{x^2}{4 q^2}) f(y)^2 \right]_{y=qx}, \n
p &\simeq& \frac{H_I^2 \tilde{m} x_1^{\frac{2}{3} \left(\frac{m}{H_I} \right)^2}}{4 \pi^2 a^2 } \int^{1/x_1}_{1/x_{\rm ini}} dq
\ q^{1+\frac{2}{3} \left(\frac{m}{H_I} \right)^2 } 
 \left[(\partial_y f(y))^2 - (\frac{1}{3}+\frac{x^2}{4 q^2}) f(y)^2 \right]_{y=qx} \ ,
\label{pgenf} \n
\eeqa
where $f(y)=\sin (y)/y.$
The kinetic terms, $(\partial_y f(y))^2 + f(y)^2$ in the square brackets,
are the same as those of the  massless case (\ref{EMTmassless}),
except for an additional power of the momentum $q$ depending on $m$. 
But since the 
kinetic terms are IR convergent, it does not affect the integrals much.
Hence, the contribution from the kinetic terms to (\ref{pgenf}) 
is given by the same form as in the massless case 
(\ref{EMTmasslessUV}). 

The contribution from the mass term, namely, 
the term proportional to $(x^2/4q^2)f(y)^2$, 
would be IR divergent if we used the massless wave function.
However, the additional power of the momentum, $q^{\frac{2}{3} \left(\frac{m}{H_I} \right)^2}$, 
which comes from the mass deformation of the wave function,
reduces the IR divergence.
In terms of $k$ and $\eta$, the mass-term contribution is written as
\beq
\rho^{\rm mass} = \frac{m^2 H_I^2}{8 \pi^2} |\eta_1|^{\frac{2}{3} \left(\frac{m}{H_I} \right)^2}
\int_{1/|\eta_{\rm ini}|}^{1/|\eta_1|} dk \ k^{-1+\frac{2}{3} \left(\frac{m}{H_I} \right)^2 }  f(k\eta)^2 \ .
\eeq
Since the function $f(k \eta)$ starts decreasing at $k \sim 1/\eta$,
it gives an effective UV cutoff, and we have
\beq
\rho^{\rm mass} \simeq \frac{3 H_I^4}{16 \pi^2} 
\left(\frac{|\eta_1|}{\eta}\right)^{\frac{2}{3} \left(\frac{m}{H_I} \right)^2}
\left[1-  \left( \frac{\eta}{|\eta_{\rm ini}|} \right)^{\frac{2}{3} \left(\frac{m}{H_I} \right)^2 }
\right] \simeq  \frac{3 H_I^4}{16 \pi^2} \left[
1-e^{-\frac{2}{3} \left(\frac{m}{H_I} \right)^2 N_{\rm eff}}
\right] \ .
\label{rhomasset}
\eeq
Here, we have used  $m \ll H_I$ and  defined an effective e-folding
\beq
N_{\rm eff} =\ln \left(\frac{|\eta_{\rm ini}|}{\eta}\right)
= N_{\rm Inf} -N_{\rm RD} \ ,
\eeq
where
\beq
N_{\rm Inf} = \ln \frac{|\eta_{\rm ini}|}{|\eta_1|} = \ln \frac{a_{\rm BB}}{ a_{\rm ini}} 
\eeq
is an e-folding number during the inflation period and $N_{\rm RD}$ is that of the RD period
defined in (\ref{N_RD}).
Thus $N_{\rm eff}$ represents  the number of modes that
were enhanced during the inflation period, and are still outside the horizon and frozen at time $\eta$.

For sufficiently large $N_{\rm eff}$, (\ref{rhomasset}) becomes $3(H_I/2 \pi)^4$. 
This is nothing but the thermal equilibrium energy at the de Sitter temperature $T=H_I/2\pi$,
and is independent of $m$. 
In contrast, (\ref{rhomasset}) can be approximated by
\beq
\rho^{\rm mass} \simeq \frac{m^2}{2} \left( \frac{H_I}{2 \pi} \right)^2 N_{\rm eff} \ ,
\label{DEsmallN}
\eeq
when the effective e-folding number satisfies 
$N_{\rm eff} < (H_I/m)^2$. For
 $m \lesssim H_0$ and $H_I \sim 10^{-5} M_{\rm Pl} $, 
this upper bound of $N_{\rm eff}$ becomes $(H_I/m)^2 \gtrsim 10^{110}$.  
Equation (\ref{DEsmallN}) is also given by a simple physical argument that 
in the de Sitter spacetime, 
an ultra-light field experiences Brownian motion
at temperature $T=H_I/2 \pi$.
The growth of fluctuation since the initial time in de Sitter space was studied 
in \cite{Linde1982,Starobinsky2,VilenkinFord,Linde2}.\footnote{See also 
\cite{Finelli1,Finelli2,Finelli3,Marozzi,Kitamoto:2010et,Kitamoto:2011yx} for recent studies.}
The pressure density is $p^{\rm mass}=-\rho^{\rm mass}$ and 
it gives a candidate for the dark energy.
As will be shown in section~\ref{sec:dark_energy},
it can explain the present dark energy 
if $N_{\rm eff}$ satisfies $N_{\rm eff}  \gtrsim 24 \pi^2 (M_{\rm Pl}/H_I)^2 \sim 10^{12}$
for $H_{\rm I}\sim10^{-5} M_{\rm Pl}$.

For a more realistic model of the inflation epoch, which is consistent with the CMB observations,
we need to consider deviations from the pure de Sitter space.
In the slow-roll approximation, or 
in cases with a nonzero but constant deceleration parameter $q\equiv -1+\epsilon\equiv -1 -\dot{H}/H^2$,
the fluctuations of the field become
\beq
\langle \phi^2 \rangle 
\propto \int_{a_{\rm ini}}^a d (\ln a)~ H^2
\eeq
(see, e.g., Ref.~\cite{Finelli2,Janssen:2008px} and H.~Kitamoto, private communication),
and the result (\ref{DEsmallN}) is modified accordingly.
However, as long as we consider a slight deviation from the de Sitter space that is required by the CMB data,
our results do not change considerably.

%%%%%%%%%%%%%%%%%%%%%%%
\subsection{Massive case at late times}
\label{sec:massive_late}
We then consider the behaviors of the EMT at late times 
 $x=\sqrt{2m/H}>\sqrt{2}$.
In this parameter region, the WKB approximation becomes valid,
and we can use the wave function (\ref{uwaveWKB}).
In performing the momentum integration for the EMT, we  divide the integration region into two:
the UV region $q >x/2$, which corresponds to $k_{\rm phy}>m$, 
and the IR region $q <x/2$, corresponding to $k_{\rm phy}<m$.

In the UV region with $q >x/2$,  the WKB wave function 
is reduced to the plane wave $\propto e^{-i q x}$ in (\ref{WKBqgtx}).
As we discussed below Eq.~(\ref{uwaveWKB}),  
the WKB wave function with the BD initial condition  
(\ref{uwaveWKB}) becomes identical with  (\ref{uwaveBB}).
Then the EMT reduces to  (\ref{pgenf}), except for the integration region,
where the lower bound of the integration region 
is replaced by $q = x/2$.
Since we are considering the UV region $q>x/2$, 
the terms  proportional to $x^2/4 q^2$  in (\ref{pgenf}) can be neglected.
Then the UV contribution to (\ref{pgenf}) becomes
\beqa
\rho^{\rm UV} &\simeq&
\frac{H_I^2 H^2}{8 \pi^2} \ln \left(\frac{2}{x x_1} \right) \simeq
\frac{H_I^2 H^2}{8 \pi^2} N_{\rm RD} \ , \n
p^{\rm UV} &\simeq& \frac{\rho^{\rm UV}}{3} \ ,
\label{rholtUV2}
\eeqa
where we have used
\beq
\ln \left(\frac{2}{x x_1} \right) = \ln \left(\frac{H_I}{m}\frac{|\eta_1|}{\eta} \right) \simeq  N_{\rm RD}
\eeq
at  $m \sim H$, or $x \sim \sqrt{2}$.
This is slightly different from $N_{\rm RD}$ defined in (\ref{N_RD}),
since the integration is cut off at $q=x/2$ $(k_{\rm phy}=m)$,
not at $q=1/x$ $(k_{\rm phy}=H)$.
However, the difference is negligible as long as we consider the region
$x \sim \sqrt{2}$, or $m \sim H$.

\begin{figure}
\begin{center}
\begin{minipage}{.45\linewidth}
\includegraphics[width=1.1 \linewidth]{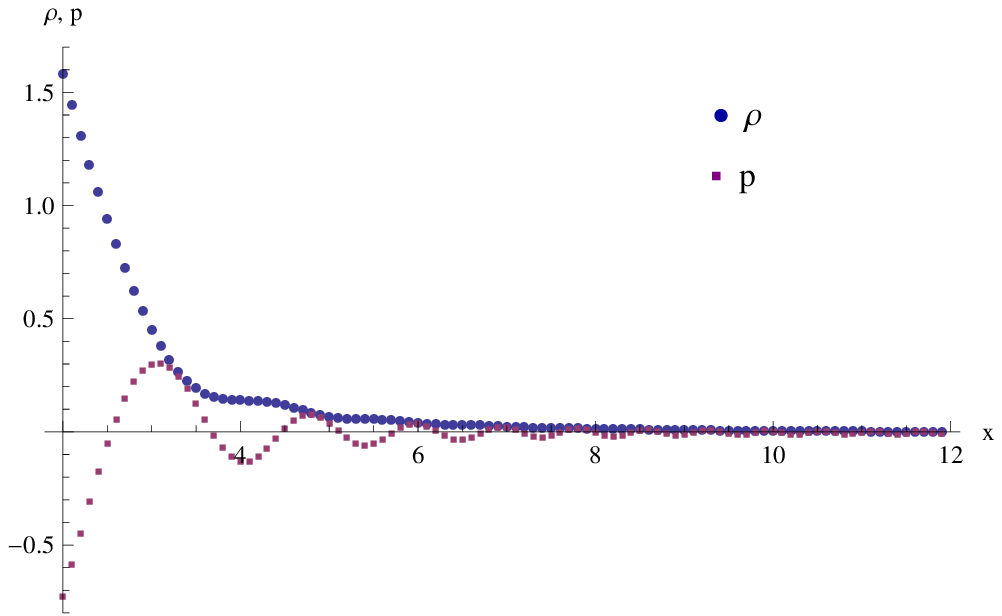}
  \end{minipage}
  \hspace{1.0pc}
\begin{minipage}{.45\linewidth}
\includegraphics[width=1.1 \linewidth]{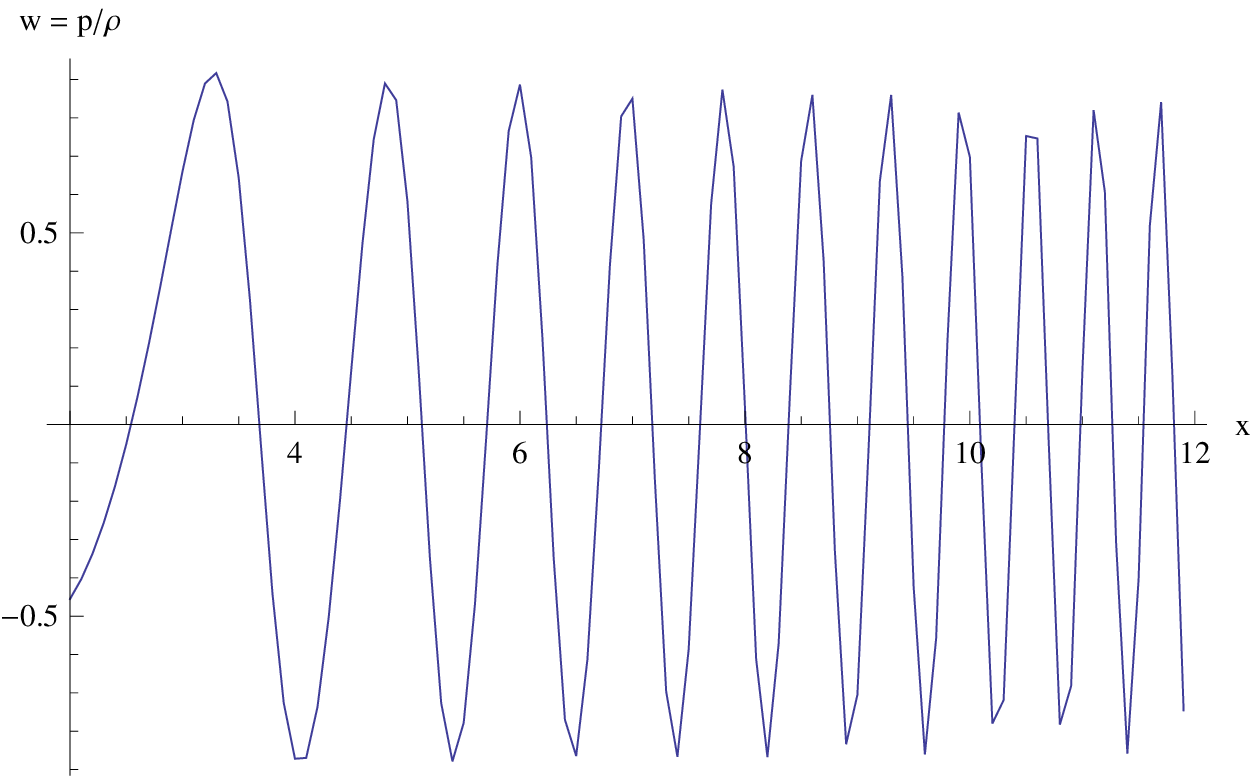}
  \end{minipage}
\begin{minipage}{.45\linewidth}
\includegraphics[width=1.1 \linewidth]{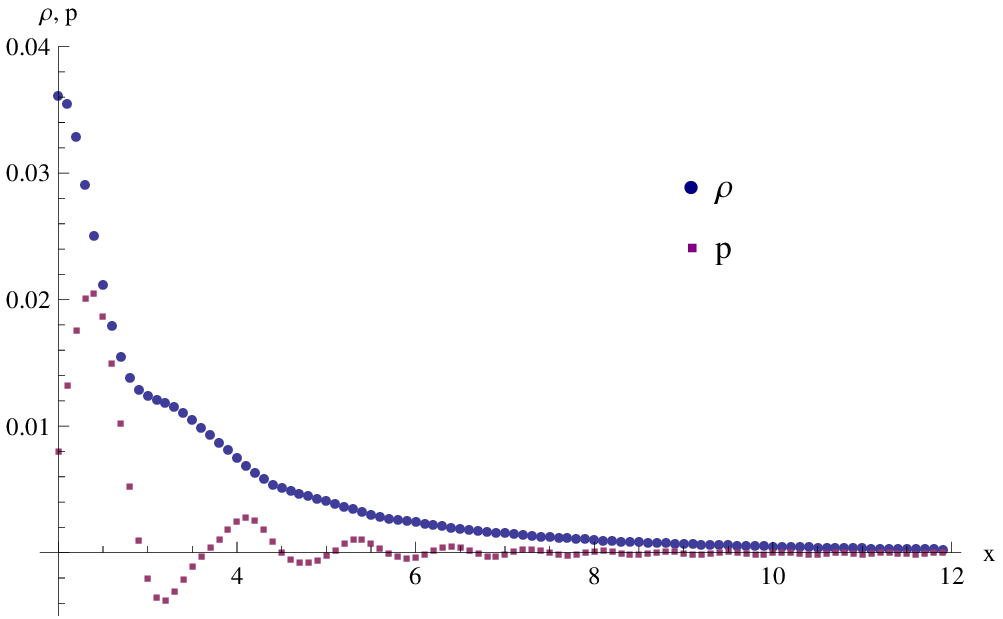}
  \end{minipage}
  \hspace{1.0pc}
\begin{minipage}{.45\linewidth}
\includegraphics[width=1.1 \linewidth]{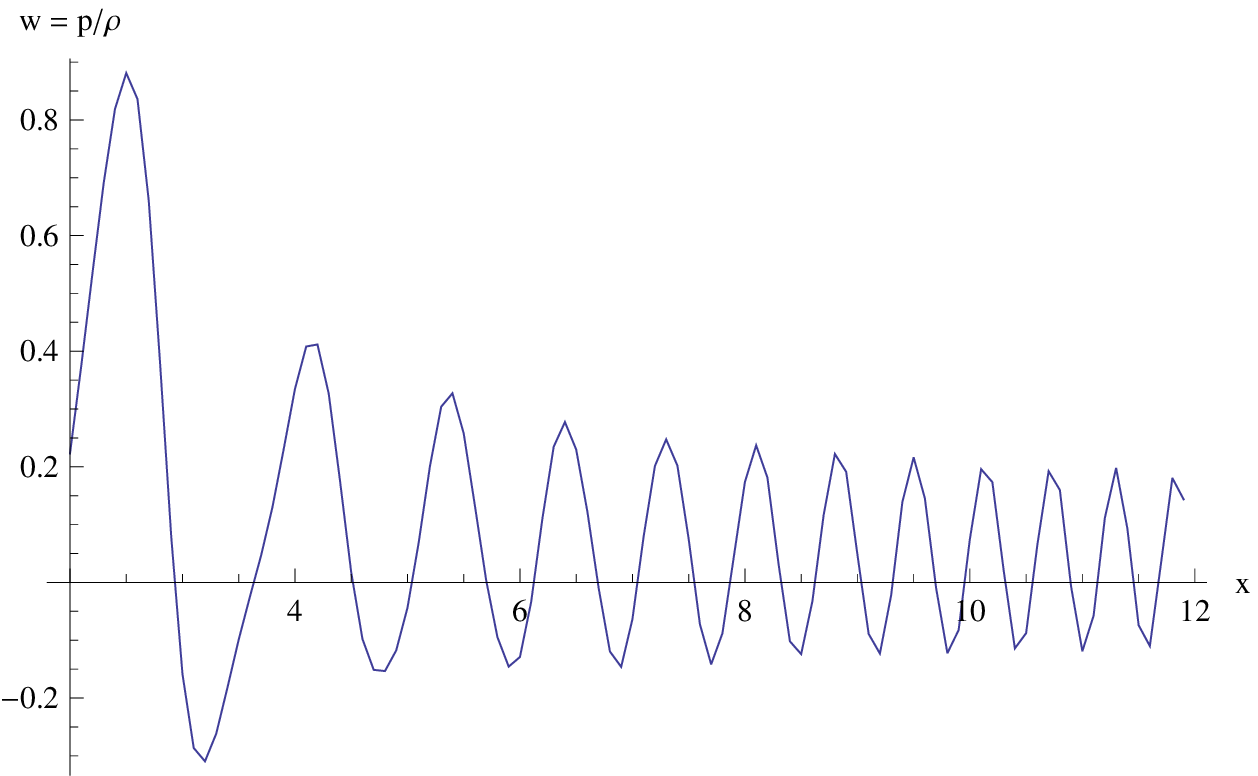}
  \end{minipage}  
\end{center}
\caption{
Time evolution of the EMT, given by the IR modes of the WKB wave function, at late times.
The horizontal axis represents time $x=(2\tilde{m})^{1/2}\eta=\sqrt{2m/H}$,
in the region $x>2$.
The left panels show $\rho^{\rm IR}$ and $p^{\rm IR}$,
normalized by $(H_I m)^2/(2\pi^2)$.
The right panels show the equation of state $w^{\rm IR}=p^{\rm IR}/\rho^{\rm IR}$.
The integration region over $q$ is taken to be $q \in [q_{\rm min}, x/2]$,
with $q_{\rm min}=0.001$ and $0.8$ 
in the upper and lower panels, respectively.
}
\label{fig:WKBlow}
\end{figure}

We next consider the IR region,  $q <x/2$ (i.e., $k_{\rm phy} <m$).
In this region,  the $q$-dependence of the WKB wave function  (\ref{uwaveWKB}) 
is given by $u \propto q^{-3/2+1/3(m/H_I)^2}$.
The other factors in  (\ref{uwaveWKB}) have mild $q$-dependences, 
and can be approximated by those at 
$q=0$.\footnote{
In this approximation, the time dependence of $\rho^{\rm IR}$ is represented by the zero-momentum mode.
In Appendix \ref{sec:IRmodes},  we give a simple derivation of the time evolution in the zero-momentum approximation. 
In Figures~\ref{fig:WKBlow} and \ref{fig:Exalow}, we 
numerically calculate the time evolution of $\rho$ without using such an approximation for the IR modes.}
Then the energy density (\ref{rhogenup}) can be evaluated as
\beqa
\rho^{\rm IR} &\simeq& \frac{H_I^2 \tilde{m} x_1^{\frac{2}{3} \left(\frac{m}{H_I} \right)^2}}{ 4\pi^2 a^2} 
\left( \frac{ \sqrt{\pi}2^{-1/4}}{\Gamma(3/4)}  \right)^2
\int^{x/2}_{1/x_{\rm ini}} dq~q^{-1+\frac{2}{3} \left(\frac{m}{H_I} \right)^2 }  \n
&&\times \left[ 4 \left( \partial_x\left(
\frac{\sin (\frac{x^2}{4}+\frac{\pi}{8} )}{x \sqrt{x}} \right)\right)^2 
+ x^2 \left(
\frac{\sin (\frac{x^2}{4}+\frac{\pi}{8} )}{x \sqrt{x}} \right)^2
\right] \ .
\label{rholtIR1} 
\eeqa
Here, we have dropped the spacial-derivative terms proportional to $q^2$ in (\ref{rhogenup})
since we are considering the IR region $q<x/2$.
Note that the $q$- and $x$-dependences of the integrand have been separated in (\ref{rholtIR1}).
The $q$-integration can be performed as 
\beqa
\int^{x/2}_{1/x_{\rm ini}} dq~q^{-1+\frac{2}{3} \left(\frac{m}{H_I} \right)^2 }
&=&  \frac{3}{2}\left(\frac{H_I}{m}\right)^2 \left(\frac{x}{2}\right)^{\frac{2}{3} \left(\frac{m}{H_I} \right)^2}
\left( 1- \left(\frac{x x_{\rm ini}}{2}\right)^{-\frac{2}{3} \left(\frac{m}{H_I} \right)^2} \right) \n
&\simeq&  \frac{3}{2}\left(\frac{H_I}{m}\right)^2
\left( 1- e^{-\frac{2}{3} \left(\frac{m}{H_I} \right)^2 N_{\rm eff}  } \right) \ ,
\eeqa
where we have used
\beq
\ln \left(\frac{x x_{\rm ini}}{2} \right) = \ln \left(\frac{m}{H_I}\frac{\eta}{|\eta_1|} \frac{|\eta_{\rm ini}|}{|\eta_1|} \right) 
\simeq  -N_{\rm RD} +N_{\rm Inf} =N_{\rm eff}
\eeq
at  $m \sim H$, or $x \sim \sqrt{2}$.

The square brakckets in (\ref{rholtIR1}) are calculated as
\beq
\Bigl[ \cdots \Bigr] = \frac{1}{x}  \left( 1 -3 x^{-2} s+\frac{9}{2}  x^{-4} (1-c) \right) 
\label{rhomassoscillation}
\eeq
where 
\beqa
s&=&\sin \left(\frac{x^2}{2}+\frac{\pi}{4}\right) =\sin \left(\frac{m}{H}+\frac{\pi}{4}\right)
=\sin \left( 2mt+\frac{\pi}{4}\right) \ , \n
c&=&\cos \left(\frac{x^2}{2}+\frac{\pi}{4}\right) =\cos \left(\frac{m}{H}+\frac{\pi}{4}\right)
=\cos \left( 2mt+\frac{\pi}{4}\right) \ .
\label{scdefinition}
\eeqa
When the higher-order corrections to the WKB approximation (\ref{WKBfunc}) are included,
the wave function $\frac{\sin(\frac{x^2}{4}+\frac{\pi}{8})}{x\sqrt{x}}$,
which lies in the parentheses in (\ref{rholtIR1}), is modified as
\beq
x^{-3/2} \left(1+{\cal O}(x^{-4}) \right)~\sin \left(\frac{x^2}{4}+\frac{\pi}{8}+{\cal O}(x^{-2}) \right) \ .
\label{WKBlm_ho}
\eeq
Then the second term with $x^{-2}$ in (\ref{rhomassoscillation}) remains unchanged,
but the third term with $x^{-4}$ receives modifications.
We thus drop it.
Therefore, (\ref{rholtIR1}) becomes 
\beq
\rho^{\rm IR} \simeq \frac{H_I^2 m^{1/2} H^{3/2}}{8\pi~\Gamma(3/4)^2}  
\left[ 1 -\frac{3}{2}\frac{H}{m} s \right] N_{\rm eff} \ .
\label{rholtIR3}
\eeq
The time dependence is dominantly given by the overall factor 
$H^{3/2} \sim a^{-3}$ with an additional oscillation given by the square-bracket factor.
Note that it is a non-analytic function of $m$ and the behavior cannot be
obtained from the massless theory by perturbation with respect to the mass. 
An interpolating solution between the early-time behavior (\ref{DEsmallN})
and the late-time behavior (\ref{rholtIR3})
can be obtained by the zero-momentum wave function given in Appendix~\ref{sec:IRmodes}.

Similarly, one can estimate the pressure density.
From (\ref{pgenup}), one obtains almost the same equation as in (\ref{rholtIR1}),
but with the second term in the square brackets
having a minus sign. Accordingly, (\ref{rhomassoscillation}) is replaced by
\beq
\frac{1}{x}  \left( c -3 x^{-2} s+\frac{9}{2}  x^{-4} (1-c) \right) \ ,
\eeq
and we have 
\beq
p^{\rm IR} \simeq \frac{H_I^2 m^{1/2} H^{3/2}}{8\pi~\Gamma(3/4)^2}  
\left[ c -\frac{3}{2}\frac{H}{m} s \right] N_{\rm eff} \ .
\label{pltIR3}
\eeq
The EMTs, (\ref{rholtIR3}) and (\ref{pltIR3}), decrease as $H^{3/2}\propto a^{-3}$,
and the equation of state oscillates as
\beq
w^{\rm IR} \simeq 
\left(c -\frac{3}{2} \frac{H}{m} s \right)/
\left(1 -\frac{3}{2} \frac{H}{m} s \right)
\simeq c - \frac{3}{2} \frac{H}{m} s(1-c) \ .
\label{wltIR}
\eeq
The behavior agrees with our knowledge that, once the 
scalar field $\phi$ starts 
oscillating in the quadratic potential $m \phi^2/2$, it behaves as a dust. 
Let us see how these behaviors are consistent with the conservation law of 
the EMT.
The energy and pressure densities behave as
\beqa
\rho^{\rm IR} &\propto& x^{-3} \left( 1 -\frac{3}{x^{2}} \sin \left(\frac{x^2}{2}+\frac{\pi}{4}\right)
+ {\cal O}(x^{-4})  \right) \ , \n
p^{\rm IR} &\propto& x^{-3} \left( \cos \left(\frac{x^2}{2}+\frac{\pi}{4}\right) 
-\frac{3}{x^{2}} \sin \left(\frac{x^2}{2}+\frac{\pi}{4}\right)+ {\cal O}(x^{-4})   \right) \ .
\eeqa
It can be easily shown that they satisfy the  conservation law of the EMT 
\beq
\partial_x \rho +3 \frac{\partial_x a}{a} (\rho+p)=0 \ 
\eeq
up to the order $x^{-4}$, where $\partial_x a/a =1/x$.
Hence, the energy density decreases as $a^{-3}$ as if it were a pressureless dust
although the pressure is nonvanishing but oscillating.
To see the consistency at the order $x^{-6}$ and higher,
we need to include higher-order corrections to the WKB approximation in (\ref{WKBfunc}). 

The ratio of the UV contribution (\ref{rholtUV2}) to 
the IR contribution (\ref{rholtIR3}) is given by 
\beq
\frac{\rho^{\rm UV}}{\rho^{\rm IR}} = \frac{\Gamma(3/4)^2}{\pi} 
 \frac{N_{\rm RD} }{N_{\rm eff} } \sqrt{ \frac{H}{m} } \ .
 \label{ratioUVIR}
\eeq
As we saw at the end of the previous section,
$N_{\rm RD} \sim 60$, and a very large e-folding $N_{\rm eff} \gtrsim 10^{12}$ is necessary. 
Note also that  $m>H$ is satisfied for the late times.
Therefore the r.h.s. in (\ref{ratioUVIR}) is much smaller than one,
and the IR contribution $\rho^{\rm IR}$ gives a dominant contribution to the EMT compared to $\rho^{\rm UV}$.

In Figure~\ref{fig:WKBlow}, 
we numerically performed the $q$-integrations of (\ref{rhogenup}) and (\ref{pgenup})
for the WKB wave function (\ref{uwaveWKB}) with (\ref{WKBfunc}), and plotted the
time evolution of the energy and pressure densities on the left.
The equation of state is plotted on the right.
The energy and pressure densities are normalized by
\beq
\frac{H_I^2 \tilde{m} (x_1)^{\frac{2}{3} \left(\frac{m}{H_I} \right)^2}}{4 \pi^2 a^2} x^2
=\frac{(x_1)^{\frac{2}{3} \left(\frac{m}{H_I} \right)^2}}{2 \pi^2}  H_I^2 m^2
\simeq \frac{1}{2 \pi^2}  H_I^2 m^2 \ .
\label{normal_fac}
\eeq
The integration region is taken as $q \in [q_{\rm min},x/2]$
to obtain the IR contribution.
The lower bound corresponds to 
\beq
q_{\rm min}=\frac{1}{x_{\rm ini}} 
\simeq e^{-N_{\rm eff}} \ \ .
\eeq
We set the lower bound of the $q$-integration to 
$q_{\rm min}= 0.001$ for the upper panels and $q_{\rm min}= 0.8$ for the lower.
The behavior of the upper panels agrees well with the analytical estimations given in this section.
The energy density decreases with $x^{-3}$ and the
equation of state oscillates between $-1$ and $1$.
By taking various values of $q_{\rm min}$ as $10^{-1}$, $10^{-2}$, $10^{-3}$, $10^{-4}$, $10^{-5}$, etc,
we have also confirmed that the magnitude of $\rho^{\rm IR}$ and  $p^{\rm IR}$ 
scales as $N_{\rm eff}=-\ln (q_{\rm min})$, as shown by (\ref{rholtIR3}) and (\ref{pltIR3}).
On the other hand, if we take the IR cutoff larger, 
the approximation that the oscillation frequency of the integrand is independent 
of the momentum in the IR region is invalidated.
Then, by summing various modes with different momenta, the oscillating behavior 
in the time direction is incoherently averaged and is expected to be diminished.
Indeed, in the lower panels, the magnitude of $p$, and, accordingly, $w$, decreases.
However, even in these large $q_{\rm min}$, the oscillating behavior of $w$ still remains.
In the real setting, as we saw before,
we need large 
$N_{\rm eff} \sim10^{12}$, and thus small $q_{\rm min} \sim  e^{-10^{12}}$ is required.
Hence, the equation of state $w$ oscillates between $w=-1$ and $1$ rather than diminishes.

%%%%%%%%%%%%%%%%%%%%%%%%%%%%%%%%%%%%
\subsection{Numerical results for the evolution of EMT} 
\label{sec:numerical}

\begin{figure}
\begin{center}
\begin{minipage}{.45\linewidth}
\includegraphics[width=1.1 \linewidth]{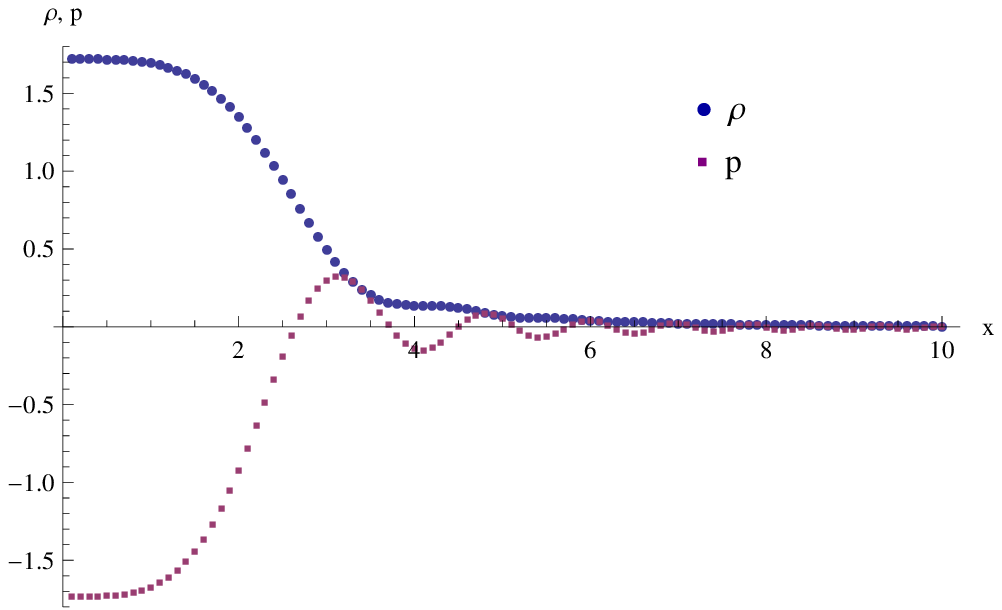}
  \end{minipage}
  \hspace{1.0pc}
\begin{minipage}{.45\linewidth}
\includegraphics[width=1.1 \linewidth]{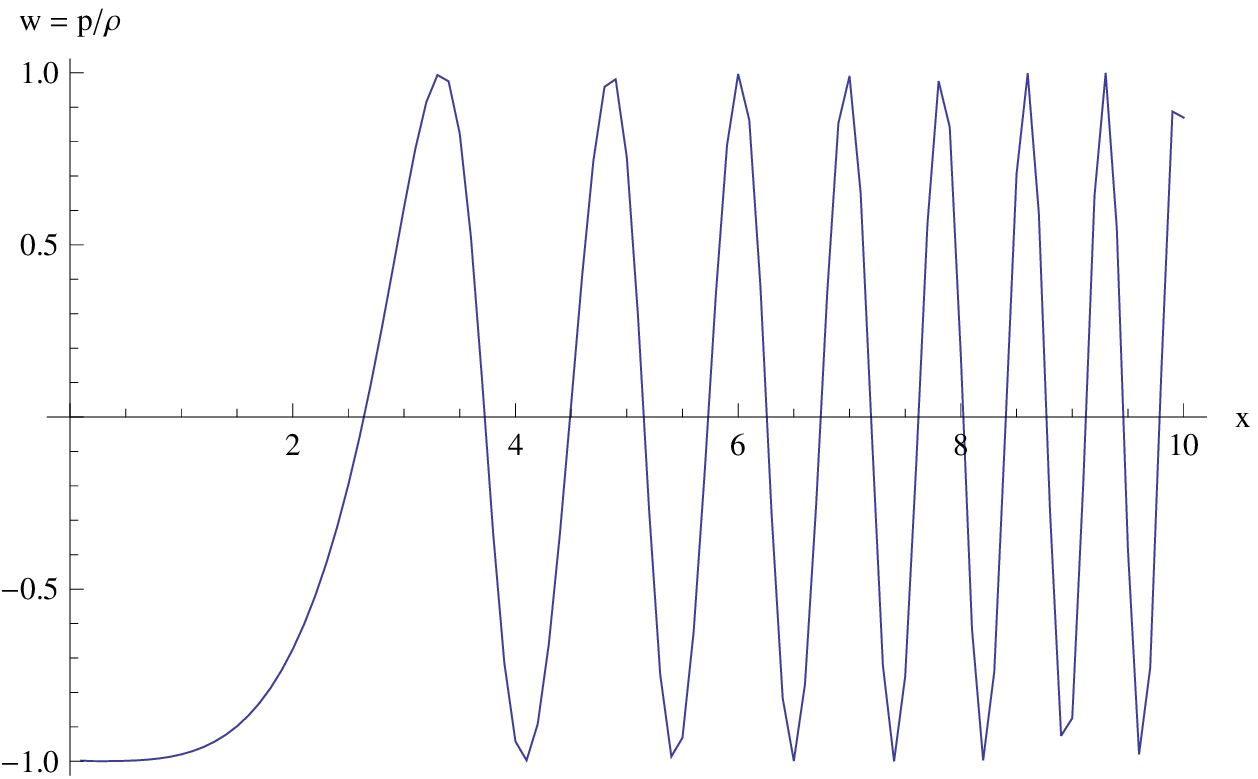}
  \end{minipage}
\end{center}  
\caption{
Similar plots to Figure~\ref{fig:WKBlow}, but using the exact wave function in  (\ref{uwavefunction}).
Note that the the horizontal axis runs from $x=0$, unlike $x=2$ in Figure~\ref{fig:WKBlow}.
The integration region over $q$ is taken to be 
$q \in [10^{-5}, 10^{-2}]$.
At early times, the IR wave function is frozen and the IR part of the EMT
behaves like the dark energy with $w=-1$. 
}
\label{fig:Exalow}
\end{figure}

In order to see how the early-  and  late-time behaviors are
 smoothly connected, we evaluate the evolution of the EMT 
by using the exact solution (\ref{uwavefunction})\footnote{{\it Exact} means that the exact wave function in the RD
period is used. For the Bogoliubov coefficient, we used an approximation (\ref{BogCapprox}). }
instead of the WKB approximation.
We insert the wave function  (\ref{uwavefunction}) into 
 the EMTs, (\ref{rhogenup}) and  (\ref{pgenup}),
and perform numerical integration over $q$. 
In order to see the contributions from the IR modes,
we set the integration region as $q \in [10^{-5}, 10^{-2}]$.
The results are shown in Figure~\ref{fig:Exalow},
where $\rho^{\rm IR}$ and $p^{\rm IR}$ are normalized by the factor (\ref{normal_fac}),
as in Figure~\ref{fig:WKBlow}.

For $x>\sqrt{2}$, the results of the exact solution in Figure~\ref{fig:Exalow} agree with 
those of the WKB wave function given in Figure~\ref{fig:WKBlow}.
For $x<1$, $w$ approaches $-1$, which agrees with the previous results given
in section~\ref{sec:massive_early}.
The early-time behavior at $x<1$ is smoothly connected with the late-time behavior at $x >\sqrt{2}$.

For technical reasons, 
the numerical integrations are performed in a restricted region of $q >10^{-5}$ and $q<10^{3}$.
In order to integrate over $q \in [q_{\rm min}, q_{\rm max}]$, where
$q_{\rm min} \sim e^{-N_{\rm eff} }< e^{-10^{12}}$
and $q_{\rm max} =x_1^{-1} =(2m/H_I)^{-1/2} \sim 10^{30}$,
we  use the analytical results based on the approximations of the wave function
discussed in the previous sections.

%%%%%%%%%%%%%%%%%%%%%%%%%%%%%%%%%%%%%%%%%%%
\section{Vacuum fluctuations as dark energy}
\label{sec:dark_energy}
\setcounter{equation}{0}

We now investigate possibilities for the vacuum fluctuations of the ultra-light scalar
to explain the dark energy at present.
We give conditions for the mass and the e-folding number, and 
then discuss how the EMT evolves through the 
RD and MD  periods.
We consider two scenarios. In  section~\ref{sec:ordinary_inflation}, 
the ordinary inflation model is discussed. In section~\ref{sec:double_inflation}, 
a double inflation model is considered, where we assume another inflation with a larger Hubble parameter
$H_P \sim M_{\rm Pl}$ before the ordinary inflation starts.

%%%%%%%%%%%%%%%%%%%%%%%%%%%%%%%%%%%%%%%%%%%%%%%%%%
\subsection{Ordinary inflation model}
\label{sec:ordinary_inflation}
We first summarize how the EMT of an ultra-light scalar field evolves in the RD period.
In the ordinary inflation model, the enhanced mode during the inflation with the largest momentum
soon enters the horizon after the inflation ends. 
Hence, the EMT is  given by a sum of UV and  IR contributions. 
At early times with $m \lesssim H$, 
as discussed in section~\ref{sec:massive_early},
the EMT is approximately given by  $\rho =\rho^{\rm UV} + \rho^{\rm IR}$
and $p=p^{\rm UV} +p^{\rm IR}$, where
\beqa
&&\rho^{\rm UV} = \frac{1}{8 \pi^2} H_I^2H^2 N_{\rm RD} \ , 
\ \  p^{\rm UV} =\frac{1}{3} \rho^{\rm UV} \ , 
\label{EMTRDearlyUV} 
\\
&&\rho^{\rm IR} = \frac{1}{8 \pi^2} H_I^2 m^2 N_{\rm eff} \ , 
\ \  p^{\rm IR} = -\rho^{\rm IR}  \ .
\label{EMTRDearlyIR}
\eeqa
The UV modes have already entered the horizon. 
$N_{\rm RD}$ 
represents the number of UV degrees of freedom.
Hence, it is time dependent.
 For the UV modes, the kinetic terms in the EMT mainly contribute and $w=1/3$ is obtained. 
The IR  modes are still out of the horizon and frozen. 
Hence, the mass term mainly contributes to the EMT, and we have $w=-1.$

If the mass of the scalar field is heavier than the Hubble parameter at the 
matter-radiation equality, $m >H_{\rm eq}=10^{-28} {\rm eV}$, 
the condition $m \gtrsim H$ becomes satisfied in the late RD period, and
the EMT is described by the late-time behaviors discussed in section~\ref{sec:massive_late}.
The coherent oscillation (the motion of the zero mode) gradually starts and 
the behavior  (\ref{EMTRDearlyIR}) of the EMT is changed to 
(\ref{rholtIR3}) and (\ref{pltIR3}):
\beq
\rho^{\rm IR} = R  
\left[ 1 -\frac{3}{2}\frac{H}{m} s \right] \ , \ \ 
p^{\rm IR} = R 
\left[ c -\frac{3}{2}\frac{H}{m} s \right]  \ ,
\label{pltIR3again}
\eeq
where $s$ and $c$ are defined in (\ref{scdefinition})
and 
\beq
R = \frac{H_I^2 m^{1/2} H^{3/2}}{8\pi~\Gamma(3/4)^2}  N_{\rm eff} \  .
\eeq
The amplitudes are proportional to $\sqrt{m}H^{3/2}$ and decay as $H^{3/2} \propto a^{-3}.$
The EMT behaves like a dust with an oscillating $w$. 
The UV contribution (\ref{EMTRDearlyUV}) remains unchanged.

\vspace{5mm}

Now let us study the evolution of the EMT in the MD period.
If the mass is lighter than $H_{\rm eq}$, 
$m <10^{-28} {\rm eV}$, the coherent oscillation of the zero mode of the scalar field
has not yet started at the beginning of  the MD period. 
In the early times of the MD period when the condition $m \lesssim H$ is satisfied, 
the EMT  is again written as a sum of the UV and IR contributions, $\rho=\rho^{\rm UV}+\rho^{\rm IR}.$
The  UV part comes from the modes that have already
entered the horizon, and is 
given by  Eq.~(7.20) of Ref.~\cite{AIS}:\footnote{In \cite{AIS}, a massless case was studied, but, 
at early times with $m<H$, the wave function is not modified much by the mass.}
\beqa
\rho^{\rm UV} = \frac{1}{8 \pi^2} H_I^2 H^2 \left( \frac{a_{\rm eq}}{a} \right) N_{\rm RD}  \ , \ \ 
p^{\rm UV} =\frac{\rho^{\rm UV}}{3} \ ,
\label{EMTMDETUV}
\eeqa
where $a_{\rm eq}$ and $a$ are the scale factor at the matter-radiation equality and at each time
 in the MD period, respectively.
$N_{\rm RD}=\ln(a_{\rm eq}/a_{\rm BB})$ is the e-folding number during the RD period and is constant in time.
Compared to the IR contributions discussed below, this UV contribution $\rho^{\rm UV}$ becomes negligible
because of the factor $(a_{\rm eq}/{a}).$
The IR part in the EMT has contributions from the kinetic term and the mass term,
$\rho^{\rm IR}=\rho^{\rm  IR,kin}+\rho^{\rm IR, mass}$.
They are evaluated as
\beqa
\rho^{\rm IR, kin}&=&\frac{3}{32\pi^2} H_I^2 H^2 \ , \ \ 
p^{\rm IR, kin}=0 \ ,
\label{kinMDET}  \\
\rho^{\rm IR, mass}&=&\frac{1}{8\pi^2} H_I^2 m^2 N_{\rm eff}\ , \ \ 
p^{\rm IR, mass}= -\rho^{\rm IR, mass} \ .
\label{massMDET}
\eeqa
$\rho^{\rm IR,kin}$ was obtained in Eq.~(7.18) in Ref.~\cite{AIS}. 
$\rho^{\rm IR,mass}$ is the same as in (\ref{EMTRDearlyIR}), but 
$N_{\rm eff}$ takes a slightly different value since it
represents the number of degrees of freedom that are still out of the cosmological horizon,
and is time dependent. 
Note that  $\rho^{\rm IR, kin}$ in (\ref{kinMDET}) receives larger contributions from 
the modes with momenta $k \sim \eta^{-1}$ (i.e., $k_{\rm phy} \sim H$). 
In contrast, $\rho^{\rm IR, mass}$ 
in (\ref{massMDET}) has  dominant contributions from  the modes with much lower momenta.
It is amusing that a single ultra-light scalar  simultaneously
contains the dark-energy-like component 
$\rho^{\rm IR,mass}$ and the dark-matter-like component $\rho^{\rm IR, kin}$.

At  later times in the MD period, when the condition $m \gtrsim H$ is satisfied,
the coherent oscillation starts and the IR contribution is changed.
As shown in Appendix~\ref{sec:IRmodes},  if time evolution is
represented by the zero-momentum approximation, 
the energy and pressure densities are given by (\ref{rhoIR_MD_ETLT}) and (\ref{pIR_MD_ETLT}):
\beqa
\rho^{\rm IR} \simeq R' \left[1-\frac{3}{2}\frac{H}{m}s\right],  \ \ \ 
p^{\rm IR} \simeq R' \left[c-\frac{3}{2}\frac{H}{m}s\right]
\label{IR_MDLT}
\eeqa
where $s$ and $c$ are defined in (\ref{scdefinitionMD}) and 
\beq
R' =  \frac{9}{32\pi^2} H_I^2 H^2 N_{\rm eff} .
\label{IR_MDLT_R}
\eeq
The amplitude of the energy density decreases as $H^2 \propto a^{-3}$.
The interpolating solution between the early-time behavior (\ref{massMDET}) and 
the late-time behavior (\ref{IR_MDLT}) is obtained 
in (\ref{rhoIR_MD_ETLT}) and (\ref{pIR_MD_ETLT}) in Appendix \ref{sec:IRmodes}.

\vspace{5mm}

A dark-energy candidate is given by (\ref{massMDET}). 
We need three conditions for (\ref{massMDET}) to explain the dark energy
in the present universe:
\beqa
\left\{  \begin{array}{ll}
 ({\rm C1}): &  m<H_0 \sim 10^{-33} \mbox{eV} \\
 ({\rm C2}): &  \rho^{\rm IR, mass}_0 > \rho^{\rm IR, kin}_0 \\
 ({\rm C3}): &  \rho^{\rm IR, mass}_0 \sim 3 M_{\rm Pl}^2 H^2_0 
\end{array}
\right . \ ,
\label{presentDEcon}
\eeqa
where $0$ denotes the present time.
The first condition states that the present time corresponds to the early times before the coherent oscillation 
of the bosonic field starts. 
The second condition requires that
 the mass-term contribution with $w=-1$ dominates over the
kinetic-term contribution with $w=0$.
This condition gives a lower bound for the mass (or a lower bound
for the effective e-folding). 
Combining these two conditions, we have
\beq
({\rm C1, C2}): \hspace{5mm}  
H_0^2 N_{\rm eff} > m^2 N_{\rm eff} > \frac{3}{4} H_0^2  \ .
\eeq
The third condtion is necessary if  the observed magnitude of the present dark energy is given by (\ref{massMDET}).
It can be written as
\beq
({\rm C3}): \hspace{5mm} 
m^2 N_{\rm eff} = 24 \pi^2 \left( \frac{M_{\rm Pl}}{H_I} \right)^2 H_0^2 \ .
\label{m2Neff}
\eeq
Inserting (C3) into (C1,C2), we have the following conditions for  $H_I$ and $N_{\rm eff}$:
\beq
 N_{\rm eff} > 24 \pi^2 \left( \frac{M_{\rm Pl}}{H_I} \right)^2 > \frac{3}{4}  \ .
 \label{Mpl_HI_Neff}
\eeq
In the ordinary inflation, we already have a constraint from the CMB observation
that $H_I < 3.6 \times 10^{-5} \ M_{\rm Pl}$. 
Then the second condition in (\ref{Mpl_HI_Neff}) is already satisfied. 
The first one requires  quite a large e-folding:
\beq
N_{\rm eff} > 1.8 \times 10^{11} \ .
\label{Neff1011}
\eeq
Similar analyses were given in \cite{Ringeval}, where
the lower bound for an e-folding number was given as $N>10^9$.
Taking numerical factors into account, we obtain a slightly different value (\ref{Neff1011}).
The result (\ref{Neff1011}) may indicate that the observed universe with the Hubble radius $1/H_0$
is embedded in a huge universe whose size is 
$e^{1.8 \times 10^{11}}$
times larger.

\vspace{5mm}

\begin{figure}[t]
\begin{center}
\begin{minipage}{.45\linewidth}
\includegraphics[width=1.1 \linewidth]{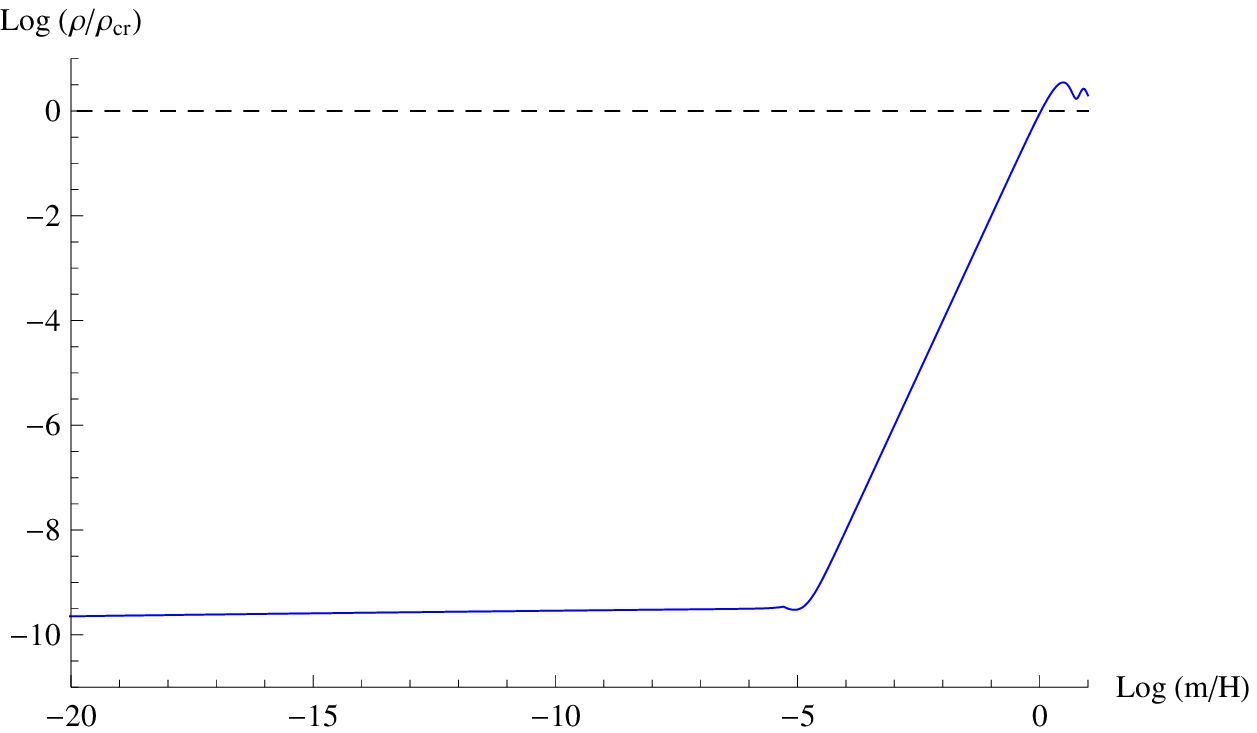}
  \end{minipage}
  \hspace{1.0pc}
\begin{minipage}{.45\linewidth}
\includegraphics[width=1.1 \linewidth]{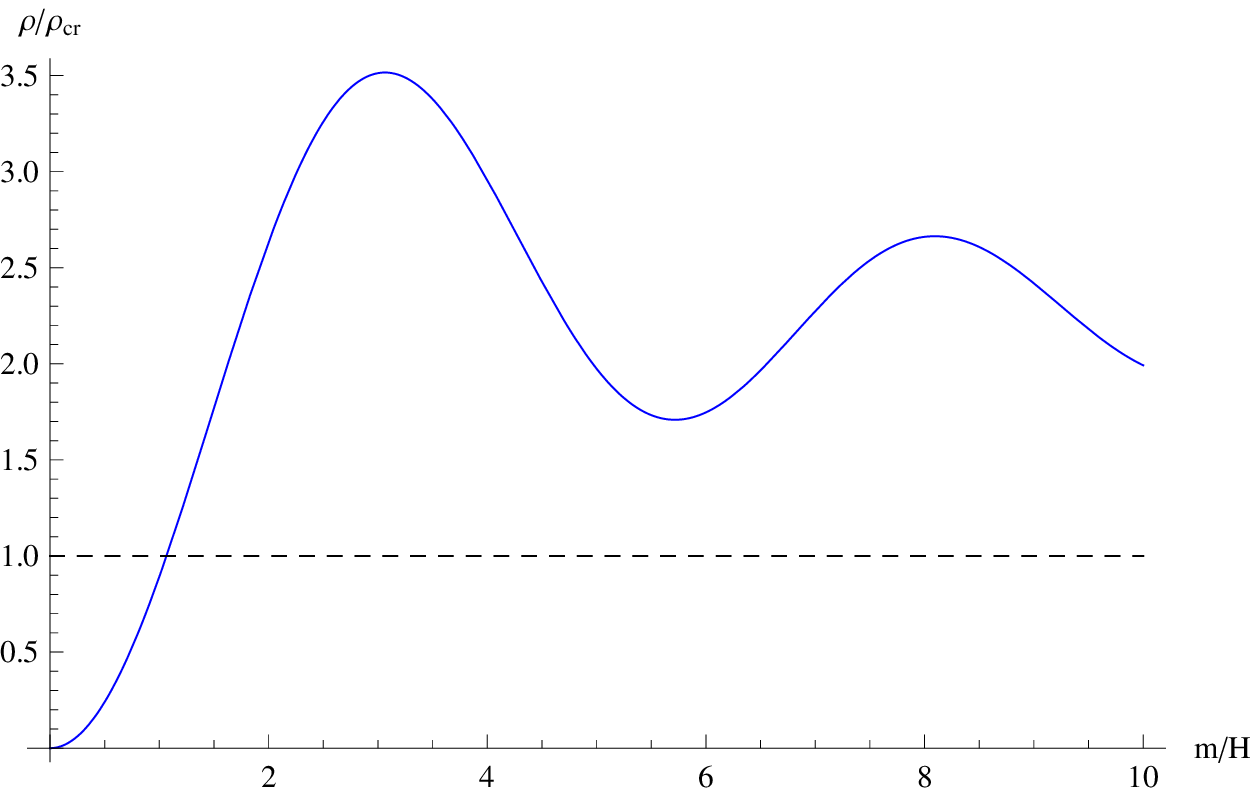}
  \end{minipage}
  \begin{minipage}{.45\linewidth}
\includegraphics[width=1.1 \linewidth]{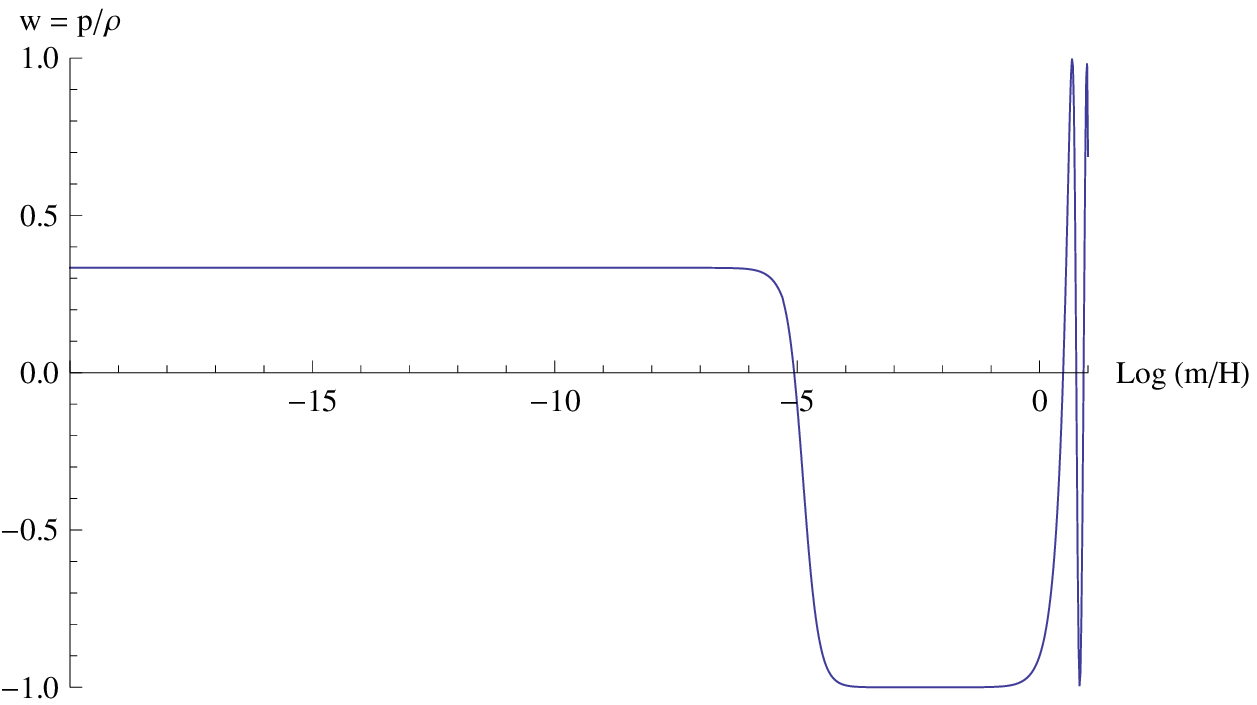}
  \end{minipage}
  \hspace{1.0pc}
\begin{minipage}{.45\linewidth}
\includegraphics[width=1.1 \linewidth]{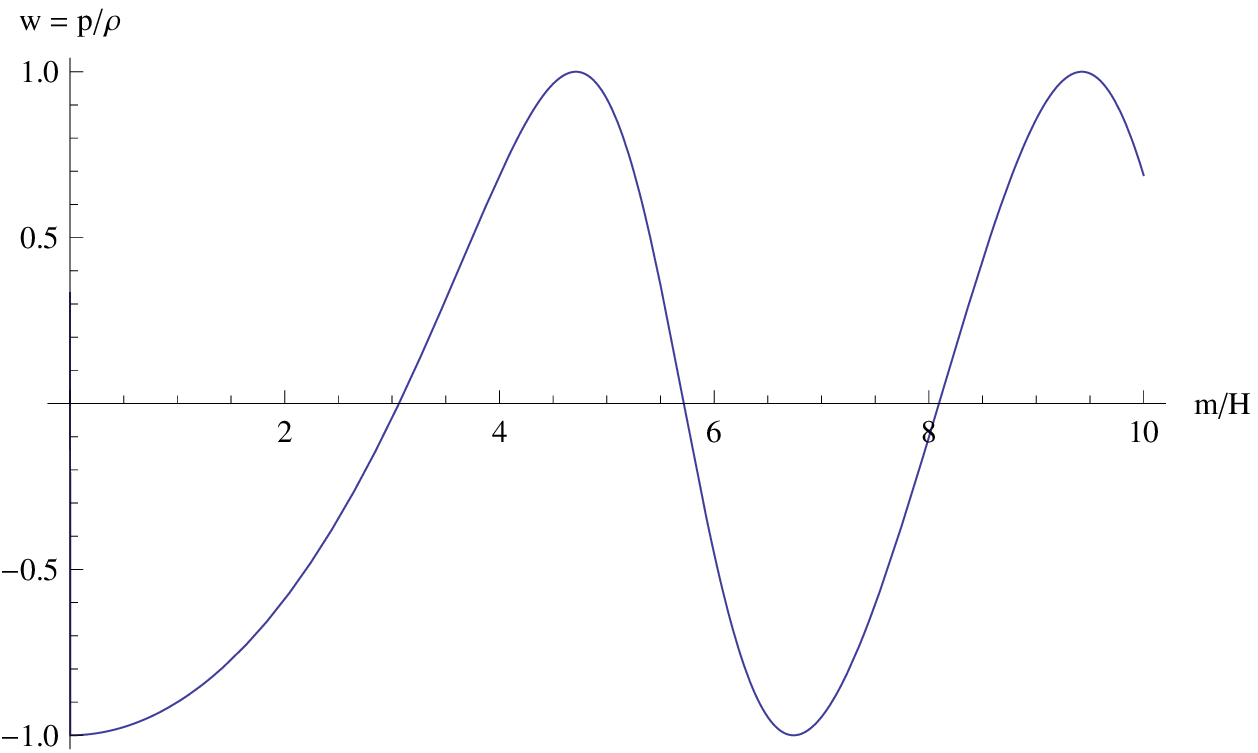}
  \end{minipage}
\end{center}  
   \caption{
Time evolution of the vacuum fluctuation generated by the ordinary inflation.
In the upper panels, the energy density $\rho$ divided by the critical value 
$\rho_{\rm cr}$ is plotted against time $m/H$.
The dashed line corresponds to the critical value.
Both horizontal and vertical axes are logarithmically scaled in the left panel. 
In the lower panels, the equation of state $w=p/\rho$ is plotted against time $m/H$.
The horizontal axis is logarithmically scaled in the left panel.
We used the following numerical values:
$H_I=3.6\times10^{-5} M_{\rm Pl}$, 
$m=H_0$, and $N_{\rm eff} = 1.8 \times 10^{11}$.
The RD period started at $m/H=1.6\times 10^{-56}$,
while the left panels display only the region $m/H>10^{-20}$.
The matter-radiation equality occurs at $m/H=5.0\times 10^{-6}$,
and the present time is $m/H=1$.
}
 \label{fig:mH_rhow}
\end{figure}

In Figure~\ref{fig:mH_rhow}, we show the time evolution of the EMT
in the RD and MD periods.
The upper panels plot the energy density $\rho$, divided by the critical value $\rho_{\rm cr}$,
while the lower panels plot the equation of state $w=p/\rho$.
We use $m/H$ for the horizontal axis to denote time evolution.
We used the following numerical values: the Planck scale
$M_{\rm Pl}=2.4 \times 10^{30}{\rm meV}$, the present Hubble parameter 
$H_{0}=1.4 \times 10^{-30}{\rm meV}$, and the redshift factor at the matter-radiation equality
$z_{\rm eq}=3.4\times 10^{3}$.
Thus the Hubble parameter at the equality is given by
$H_{\rm eq}=H_0 z_{\rm eq}^{3/2}=2.8\times10^{-25}{\rm meV}$.
In drawing the figures, we chose the Hubble parameter during the inflation at
$H_I=3.6\times10^{-5} M_{\rm Pl}=8.6 \times 10^{25}{\rm meV}$, which is 
the upper bound of $H_I$ from the CMB constraint.
The mass of the ultra-light scalar field is chosen at $m=H_0$.
Then condition (C3) in (\ref{m2Neff}) requires that $N_{\rm eff} = 1.8 \times 10^{11}$.
For the above parameters, 
 the RD period started at $m/H=H_0/H_I=1.6\times 10^{-56}$,
the matter-radiation equality occurs at $m/H=H_0/H_{\rm eq}=5.0\times 10^{-6}$,
and the present time corresponds to $m/H=H_0/H_0=1$.

At early times, the UV contribution to the kinetic term $\rho^{\rm UV}$ is dominant and gives 
the equation of state $w=1/3$, while 
its magnitude is much smaller than the critical value.
As time passes, the IR contribution to the mass term $\rho^{\rm IR,mass}$ 
grows and dominates over $\rho^{\rm UV}$ at some time, when $w$ changes to $-1$.
For the parameters we chose,
the transition  between the above two behaviors occurs, accidentally, at around  the same time as 
the matter-radiation equality.\footnote{
Comparing (\ref{EMTRDearlyUV}) and  (\ref{EMTRDearlyIR}) with (\ref{m2Neff}), 
we find that the time $m/H$ of this transition is proportional to  $m H_I$.
Then, if we choose a smaller $H_I$, the transition occurs earlier.
We also note that the period with $w=0$ might
appear between the eras of $w=1/3$ and $w=-1$, when $\rho^{\rm IR,kin}$ in (\ref{kinMDET}) is dominant.
However,  (\ref{kinMDET}) could dominate (\ref{massMDET}) with  (\ref{m2Neff})
when $(H/H_0)^2>2.4 \times 10^{11}$, which is actually larger than $(H_{\rm eq}/H_0)^2=3.9 \times 10^{10}$.
Since (\ref{kinMDET}) appears only in the MD period,
the intermediate stage with $w=0$ never arises for any possible values of the parameters.
}

As time passes further, the energy density approaches the critical value,
which gives the present dark energy,
and then $w$ begins to oscillate.
As the lower panels in Figure~\ref{fig:mH_rhow} show, 
the oscillating behavior already begins at present $m/H=1$
and the present equation of state is  given by $w\sim -0.9$.
If we choose a smaller mass like $m\sim 0.1 H_0$,
the equation of state $w=-1$ can be realized at present, 
but  we need a 100 times larger $N_{\rm eff}$.
Such difference will be detected in future observations.

It is also interesting to note that the energy density goes over the critical value when $m>H$,
as shown in the upper panels in Figure~\ref{fig:mH_rhow}.
Indeed, the coefficient of (\ref{IR_MDLT_R}) is $9/4$ times larger than that
of  (\ref{massMDET}).
Then condition (C3) in (\ref{presentDEcon}) necessarily 
leads to a situation in which the energy density exceeds the critical density of the background universe in the future.
So we need to take  into account the back reactions to the geometry to extrapolate our analysis
 to obtain the behaviors of the future universe.
 
In \cite{Habara:2014sha}, it was shown that the vacuum energy of a quantum field 
drives de Sitter expansion in the inflation period,
by studying the back reaction in a self-consistent way. 
It is also an interesting theoretical problem to 
study the late-time behavior to understand the fate of the universe. 
In this case, the interplay between the scale factor and the IR behavior of the  wave function,
with the Bunch-Davies initial condition, 
determines the dynamics of the universe self-consistently.

%%%%%%%%%%%%%%%%%%%%%%%%%%%%%%%%%%%%%%%%%%%%%%%%%%
\subsection{Double inflation model}
\label{sec:double_inflation}
Let us now consider another possibility that the EMT of the ultra-light scalar field is enhanced due to the 
fluctuations created before the ordinary inflation period.
As a simple example, we consider a cosmic model with two inflationary periods:
the ordinary inflation with the Hubble parameter $H_I$ and a {\it pre-inflation}
with a larger $H_P$ before the ordinary inflation.
A similar model was studied to obtain a modified CMB spectrum (see, e.g., \cite{Silk:1986vc}).
It will be reasonable that such a period existed before the ordinary inflation, 
because, in the very early time of the Planck scale,
quantum gravity effects possibly  generated a large-Hubble de Sitter expansion. 
Some concrete examples are  the Starobinsky type of inflation \cite{Starobinsky} 
and the eternal inflation \cite{CDL,HM,Linde,FSSY}, where our universe is  surrounded by a region with a
larger Hubble parameter. 

In addition to the Hubble parameter in the pre-inflation period $H_P$,
the double inflation model has another important parameter, i.e.,
 the conformal time $\eta_*$ when the pre-inflation period ended and the ordinary inflation  started. 
As studied in Ref.~\cite{AIS},
the wave function is enhanced to a larger amplitude of the order of $H_P$ during the pre-inflation period,
and the enhanced modes are  restricted within the momentum region $k < 1/|\eta_{*}|$.
If $|\eta_*|$ is larger than the current conformal time $\eta_0$,
all the enhanced modes are still outside the current horizon, and do not affect the CMB data.
In such a case, $H_P$ is not constrained by the CMB observation,
and can be taken as large as the Planck scale $M_{\rm Pl}$.\footnote{
In  Ref.~\cite{AIS}, we have also studied the intermediate stage between the pre-inflation
and inflation periods. Since the relevant modes are outside the horizon at the intermediate stage,
our results are not affected much by this stage, such as by the reheating processes after pre-inflation.}

\vspace{5mm}
We now study the time evolution of the EMT.
We first consider the RD period. 
The EMT is given by a sum $\rho=\rho^{\rm UV} + \rho^{\rm IR}$.
Here $\rho^{\rm UV}$ is the contribution to $\rho$ from the UV modes that are enhanced during the ordinary
inflation, but not during the pre-inflation. 
Hence, it is given by the same equation as (\ref{EMTRDearlyUV}):
\beq
\rho^{\rm UV} = \frac{1}{8 \pi^2} H_I^2H^2 N_{\rm RD} \ , 
\ \  p^{\rm UV} =\frac{1}{3} \rho^{\rm UV} \ .
\label{EMTUV_PI}
\eeq
On the other hand, the IR part  $\rho^{\rm IR}$ is the contribution from the IR modes
that are greatly enhanced during the pre-inflation with  $H_P$.
$\rho^{\rm IR}$ is given by
the sum of the mass and the kinetic terms $\rho^{\rm IR}=\rho^{\rm IR,mass} + \rho^{\rm IR, kin}.$
Before the coherent oscillation starts, i.e., when $m <H$, 
the mass term in the EMT becomes
\beq
\rho^{\rm IR, mass} = \frac{1}{8 \pi^2} H_P^2 m^2 N_{\rm Preinf} \ , 
\ \  p^{\rm IR, mass} = -\rho^{\rm IR, mass}  \ ,
\label{IRmass_RD_ET_PI}
\eeq
where
\beq
N_{\rm Preinf}=\ln \left| \frac{\eta_{\rm ini}}{\eta_*} \right|
\eeq
is an e-folding number during the pre-inflation period.
Here, $\eta_{\rm ini}$ and $\eta_*$ denote the conformal time at 
the beginning and end of the pre-inflation period.
The kinetic term in the EMT becomes (see Eq.~(8.22) in Ref.~\cite{AIS})
\beqa
\rho^{\rm IR, kin} &=& 
\frac{H_P^2}{32\pi^2 \eta_*^2 a^2}
=\frac{H_P^2 H_0^2}{128\pi^2}  \left( \frac{\eta_0}{\eta_*} \right)^2  \left( \frac{a_0}{a} \right)^2 
=\frac{H_P^2 H_0^2}{128\pi^2}  \left( \frac{\eta_0}{\eta_*} \right)^2  z_{\rm eq}^2 \frac{H}{H_{\rm eq}}
\ , \n
p^{\rm IR, kin} &=& -\frac{1}{3} \rho^{\rm IR, kin} \ .
\label{IRkin_RD_ET_PI}
\eeqa
As mentioned above, we assume that 
all the enhanced modes are out of the current horizon.
Thus, they are also always out of horizon and frozen in the past.
Hence, the time-derivative term in the EMT vanishes,
and the spacial-derivative term gives (\ref{IRkin_RD_ET_PI}) with $w=-1/3$.

\vspace{5mm}

In the MD period, the UV contribution becomes negligibly small as in (\ref{EMTMDETUV}).
At early times ($m \lesssim H$), from the IR contributions, the mass term in the EMT becomes
\beq
\rho^{\rm IR, mass} = \frac{1}{8 \pi^2} H_P^2 m^2 N_{\rm Preinf} \ , 
\ \  p^{\rm IR, mass} = -\rho^{\rm IR, mass}  \ .
\label{IRmass_MD_ET_PI}
\eeq
At $\eta <|\eta_*|$, the kinetic term in the EMT becomes
\beq
\rho^{\rm IR, kin} =
\frac{H_P^2 H_0^2}{128\pi^2}  \left( \frac{\eta_0}{\eta_*} \right)^2  \left(\frac{H}{H_{0}}\right)^{4/3}
\ , \ \ 
p^{\rm IR, kin} = -\frac{1}{3} \rho^{\rm IR, kin}  \ ,
\label{IRkin_MD_ET_PI}
\eeq
as in (\ref{IRkin_RD_ET_PI}). 
On the other hand, at $\eta >|\eta_*|$, some of the enhanced modes have entered the current horizon.
Then, as in (\ref{kinMDET}), we have
\beq
\rho^{\rm IR, kin} =
\frac{3}{32\pi^2} H_P^2 H^2
\ , \ \ 
p^{\rm IR, kin} = 0  \ .
\label{IRkin_MD_ET2_PI}
\eeq
At late times ($m \gtrsim H$), the IR contribution becomes (\ref{IR_MDLT}) and (\ref{IR_MDLT_R}),
 with $H_I$ and $N_{\rm eff}$ replaced by $H_P$ and $N_{\rm Preinf}$.
The oscillating behavior of $w$ is obtained again.

\vspace{5mm}

As in the ordinary inflation case, (\ref{IRmass_MD_ET_PI}) gives a dark-energy candidate.
In order to explain the dark energy in the present universe, the three conditions (\ref{presentDEcon})
are required.
The first and second conditions give
\beq
({\rm C1, C2}): \hspace{5mm}  
 H_0^2 N_{\rm Preinf} > m^2 N_{\rm Preinf} > \frac{1}{16} H_0^2 \left( \frac{\eta_0}{\eta_*} \right)^2 \ ,
\eeq
while the third one becomes
\beq
({\rm C3}): \hspace{5mm} 
m^2 N_{\rm Preinf} = 24 \pi^2 \left( \frac{M_{\rm Pl}}{H_P} \right)^2 H_0^2 \ .
\label{m2Npreinf}
\eeq
Inserting (C3) into (C1, C2), we have
\beq
 N_{\rm Preinf} > 24 \pi^2 \left( \frac{M_{\rm Pl}}{H_P} \right)^2 >\frac{1}{16} \left( \frac{\eta_0}{\eta_*} \right)^2  \ .
 \label{Mpl_HI_Npreinf}
\eeq
Since the enhanced modes must be outside the horizon in the present universe, 
$|\eta_*| > \eta_0/(2\pi)$ needs to be satisfied,
which gives $(1/16) (\eta_0/\eta_*)^2< \pi^2/4$.
Then, a large Hubble parameter as $H_P \sim M_{\rm pl}$ satisfies the second inequality in (\ref{Mpl_HI_Npreinf}).
The first one requires that the e-folding number during the pre-inflation must satisfy
\beq
N_{\rm Preinf} > 24 \pi^2 \sim 2.4 \times 10^{2}.
\eeq
Compared to the ordinary inflation, the e-folding number does not need to be as large
as (\ref{Neff1011}).

\vspace{5mm}

\begin{figure}[t]
\begin{center}
\begin{minipage}{.45\linewidth}
\includegraphics[width=1.1 \linewidth]{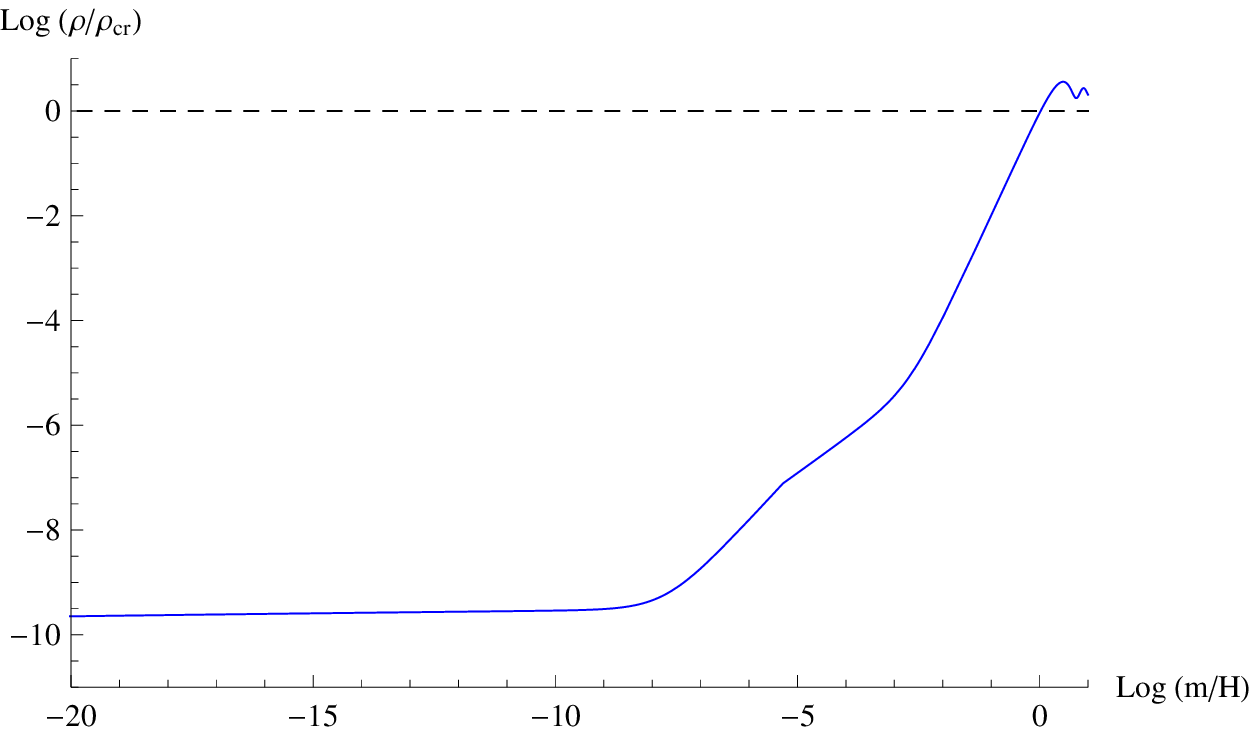}
  \end{minipage}
  \hspace{1.0pc}
\begin{minipage}{.45\linewidth}
\includegraphics[width=1.1 \linewidth]{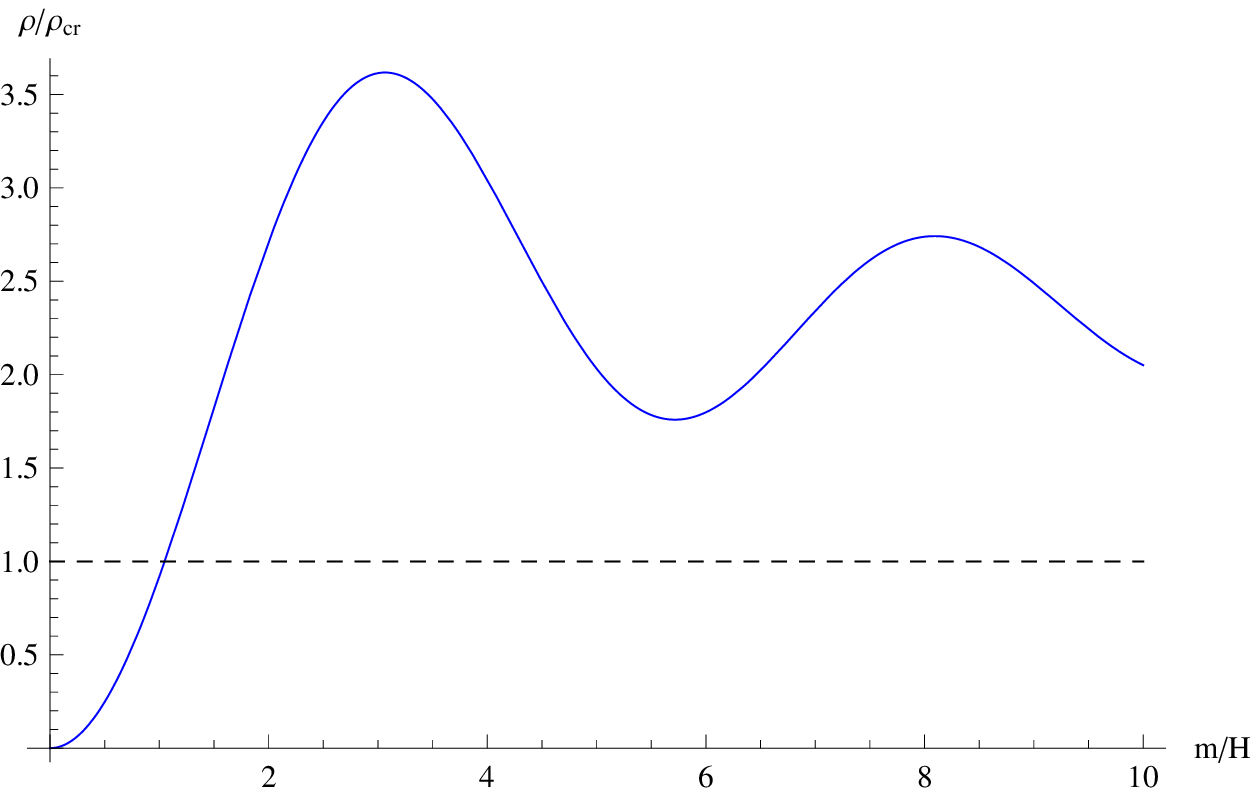}
  \end{minipage}
\begin{minipage}{.45\linewidth}
\includegraphics[width=1.1 \linewidth]{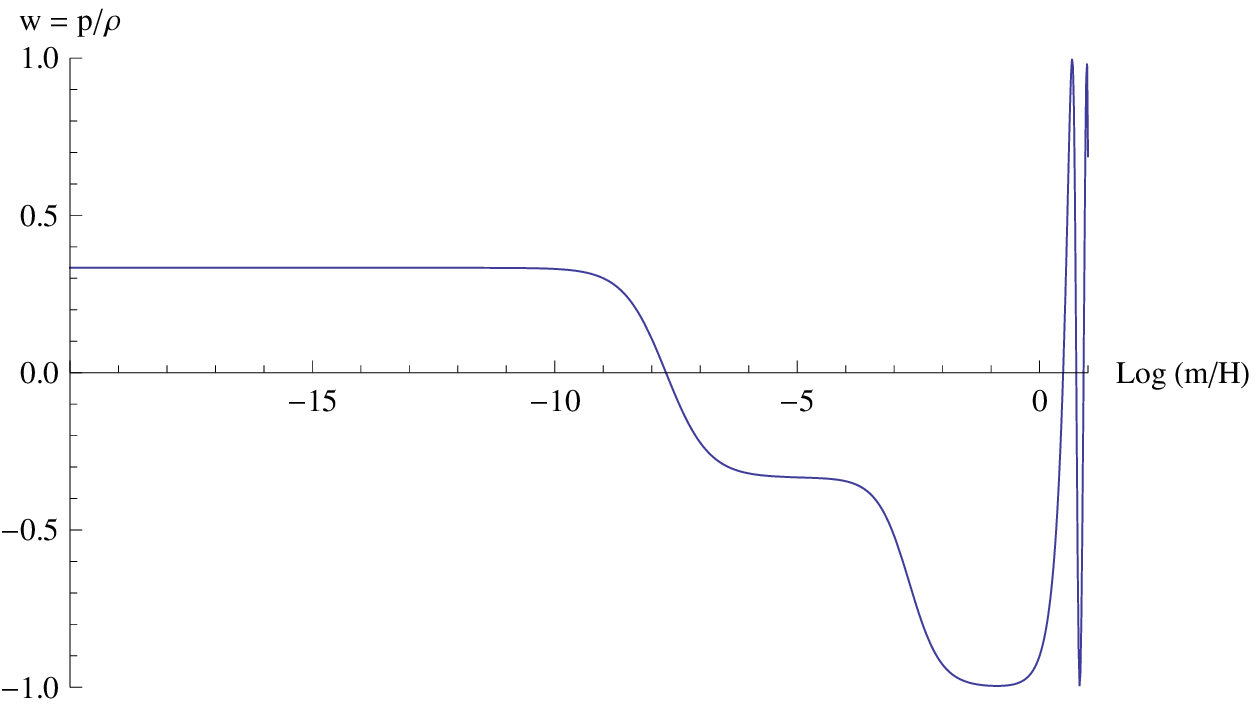}
  \end{minipage}
  \hspace{1.0pc}
\begin{minipage}{.45\linewidth}
\includegraphics[width=1.1 \linewidth]{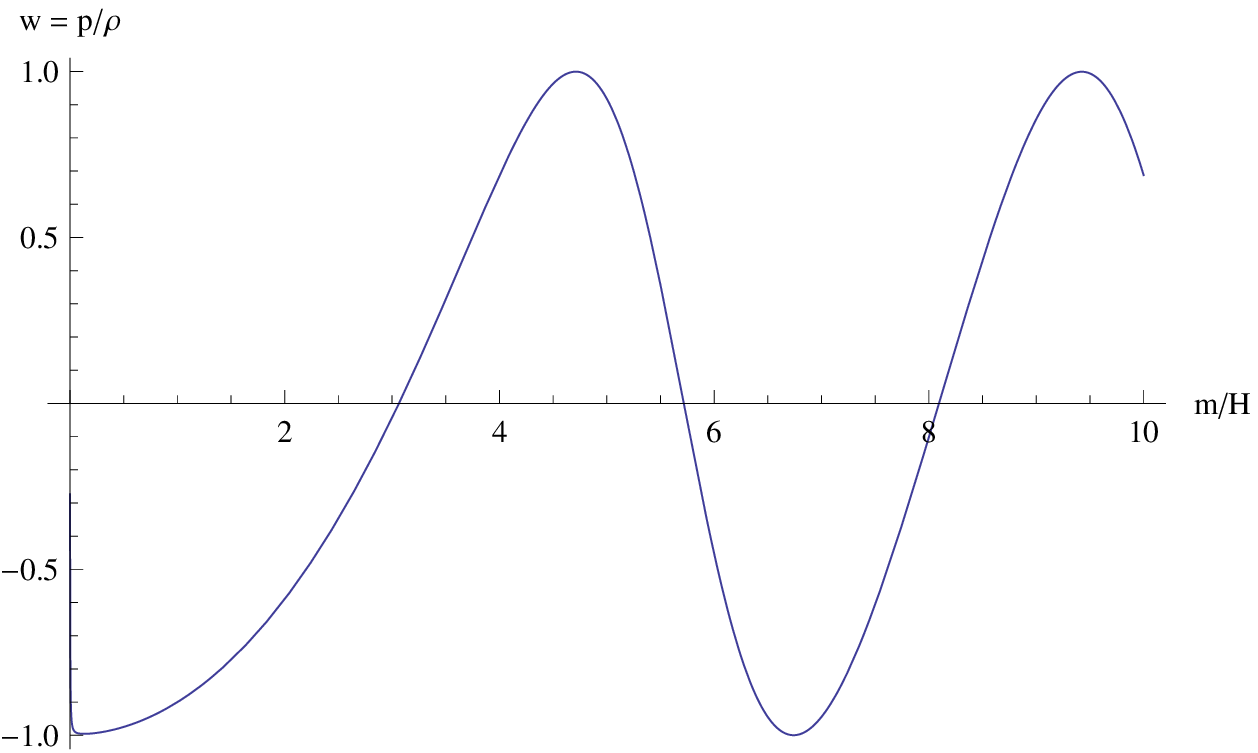}
  \end{minipage}  
\end{center} 
   \caption{
Similar plots to Figure~\ref{fig:mH_rhow}, but for the vacuum fluctuations generated by 
the pre-inflation.
The same numerical values are used as in Figure~\ref{fig:mH_rhow}.
For the additional parameters, we used $H_P=M_{\rm Pl}$, $|\eta_*|=\eta_0$,
and thus $N_{\rm Preinf}=2.4 \times 10^{2}$.
Compared to Figure~\ref{fig:mH_rhow}, 
a new state with $w=-1/3$ appears 
during the transition from $w=1/3$ to $w=-1$.
}
 \label{fig:PI_mH_rhow}
\end{figure}

In Figure~\ref{fig:PI_mH_rhow}, we show the time evolution 
of the energy density $\rho/\rho_{\rm cr}$
and the equation of state $w=p/\rho$
in the RD and MD periods.
We take the same parameters as in Figure~\ref{fig:mH_rhow}.
For the additional parameters, we use $H_P=M_{\rm Pl}$ and $|\eta_*|=\eta_0$.
Then $N_{\rm Preinf}=2.4 \times 10^{2}$ is required by (\ref{m2Npreinf}).

At early times, the UV contribution to the kinetic term $\rho^{\rm UV}$ is dominant and gives $w=1/3$, while 
its magnitude is much smaller than the critical value.
As time passes, the IR contribution to the kinetic term $\rho^{\rm IR,kin}$ dominates, 
and the era with $w=-1/3$ starts.
As time passes further, the IR contribution to the mass term $\rho^{\rm IR,mass}$ dominates, 
and the era with $w=-1$ starts.
At later times, when $m \gtrsim H$, the era with oscillating $w$ starts.

The existence of the intermediate stage with $w=-1/3$ depends on 
the values of the free parameters that we take.
The transition time $m/H$ from $w=1/3$ to $w=-1/3$, i.e.,
when $\rho^{\rm IR, kin}$  dominates $\rho^{\rm UV}$, can be shown to be proportional to
$m(H_I \eta_*/H_P)^2$, by comparing (\ref{EMTUV_PI}) and (\ref{IRkin_RD_ET_PI}).
On the other hand, the time $m/H$
when $\rho^{\rm IR, mass}$ dominates over $\rho^{\rm UV}$ is proportional to $m H_I$,
by comparing (\ref{EMTUV_PI}) and (\ref{IRmass_RD_ET_PI}) 
with (\ref{m2Npreinf}).
Hence, if we choose a larger $\eta_*$ and/or a smaller $H_P$,
the era with $w=-1$ starts before $w=-1/3$ might start,
and thus the intermediate stage with $w=-1/3$ does not arise.

%%%%%%%%%%%%%%%%%%%
\section{Conclusions and discussions}
\label{sec:concl}
\setcounter{equation}{0}

In this paper, we have calculated the time evolution of the energy-momentum tensor of 
an ultra-light scalar field with a mass $m \lesssim 10^{-33} \mbox{eV}$.
In the case of axion-like particles, the initial condition is set by hand by the misalignment 
mechanism. We instead assume that the fluctuations generated during de Sitter expansion
in the primordial inflation
gave the initial condition of the amplitude of the vacuum energy. 
If the fluctuations are enhanced during
 the ordinary inflation with the Hubble parameter $H_I \sim 10^{-5} M_{\rm Pl}$,
a very large e-folding $N_{\rm eff} \sim 10^{12}$ is necessary to explain the dark energy 
at present.  But, if we consider a cosmic history
with another Planckian universe with a large Hubble parameter $H_P \sim M_{\rm Pl} $ before the
ordinary inflation, a much smaller e-folding number $N_{\rm eff} \sim 240$ during the pre-inflation
is sufficient.
We furthermore calculated how the dark energy evolves in future,
though the back reaction to the geometry becomes relevant and needs to be taken 
in a self-consistent manner.
The amplitude decreases as $a^{-3}$ where $a$ is the scale factor, like a dust,
and the equation of state oscillates between $w=-1$ and $1$ with a large oscillation period $1/m$.
If the mass is a bit larger than the current Hubble, e.g., $m=10^{-32} \mbox{eV}$, 
such an oscillatory behavior may be detectable
(see, e.g., \cite{Marsh:2014xoa}).
In most studies of the quintessence scenario, the classical equation for 
the zero mode is used to compare with the observational data. 
It is interesting to extend the analyses to include the nonzero-momentum modes discussed in this paper
and to give more detailed 
observational constraints on  the model parameters.

Another important issue not discussed in the present paper is the effect of interactions. 
In the de Sitter expanding universe, we often encounter IR divergences $\ln (k |\eta|)$
when we calculate loop corrections of various quantities (see, e.g., \cite{Tanaka}).
The IR divergences are related to the secular time growth of 
these quantities in the $|\eta| \rightarrow 0$ limit and in many cases they can be resummed. 
For example, the secular growth in a massless scalar  with  $\lambda \phi^4$ 
interactions can be cured by resumming the logarithmic factors so that 
the massless field  acquires an effective mass $m_{\rm eff}^2 = \lambda/2 (H_I/2\pi)^2 N$,
where $N$ is an e-folding number \cite{StaYoko,Burgess}.  
Then such an interaction generates the vacuum energy proportional to 
$\rho \sim \lambda/8 (H_I/2\pi)^4 N^2$. For an infinite $N$, it approaches an 
equilibrium value $3 H_I^4 /16\pi^2$ that is independent of $\lambda$.
The ratio of this energy density 
to  the critical density of the present universe is given by
$\Omega_{\phi^4} \propto \lambda  N^2 H_I^4/(384 \pi^4 M_{\rm Pl}^2 H_0^2)$.
If $H_I =10^{-5} M_{\rm Pl}$ and $N=10^2$, it becomes 
$\Omega_{\phi^4} \sim \lambda \times 10^{100}.$
Hence, unless we take a very small coupling $\lambda \sim 10^{-100}$, it exceeds the critical energy density of the universe. 
We want to come back to this issue in the future.

\section*{Acknowledgements}
The authors would like to thank Yasuhiro Sekino for insightful discussions and a collaboration
at the early stage of the investigations,
and Hiroyuki Kitamoto for useful discussions and comments. 
This work is supported
in part by a Grant-in-Aid for Scientific Research
(Nos. 23244057, 23540329, and 24540279) from 
the Japan Society for the Promotion of Science.
 This work is also partially supported by
``The Center for the Promotion of Integrated Sciences (CPIS)'' of Sokendai.

\appendix

%%%%%%%%%%%%%%%%%%%%%%%%%%%%%%%%%%%%%%%%%%%%%%%%%%%%%%%%%%%%%%
\section{Time evolution of the zero-momentum mode}
\label{sec:IRmodes}
\setcounter{equation}{0}

In this appendix, we study zero-momentum modes and calculate the EMT by using them.
We obtain the time evolution of the EMT in the MD period, 
which interpolates the early-time (\ref{massMDET}) and late-time behaviors (\ref{IR_MDLT}). 
We also reproduce the time evolution in the RD period,
which interpolates (\ref{DEsmallN}) and (\ref{rholtIR3}).\footnote{The time evolution of zero-momentum wave function 
is also investigated in \cite{Marsh} to discuss
how an ultra-light scalar affects the growth rate of cosmological perturbation.}

The wave equation, $(\Box +m^2)u=0$, in the Robertson-Walker metric is written as
\beq
u''+2{\cal H} u'+\left(k^2+(ma)^2 \right)u=0 \ ,
\eeq
where $'  = \partial_\eta$, ${\cal H}=a'/a$, and $\eta$ is the conformal time.
In terms of the physical time $t$, it is written as
\beq
\ddot{u}+3H \dot{u}+\left((k/a)^2+m^2 \right)u=0 \ ,
\label{eq_u_t}
\eeq
with $\dot{}  = \partial_t$ and $H=\dot{a}/a$.
For sufficiently low-momentum modes, 
we can neglect the $(k_{\rm phy})^2=(k/a)^2$ term in (\ref{eq_u_t}) and approximate the equation as
\beq
\ddot{u}+3H \dot{u}+ m^2 u=0 \ .
\label{eq_u_t_IR}
\eeq

We first see the freezing behavior of the low-momentum wave function at early times
($m<H$) in the RD and MD eras. 
Then, we can neglect the  $m^2$ term in (\ref{eq_u_t_IR}), 
which is easily solved as
\beq
u=F +G \int dt~a^{-3} \ ,
\label{ufrozen}
\eeq
where $F$ and $G$ are arbitrary constants.
The first term with $F$ is time independent and represents the frozen wave function.
Imposing the Bunch-Davies initial condition,
$F \neq 0$ and $G=0$ are chosen.
Once these coefficients are fixed by the initial condition, the solution (\ref{ufrozen}) continues to be valid 
throughout the history, either in the RD or MD periods,
as long as the condition $k_{\rm phy}, m \ll H$ is satisfied.

We next consider the case where
the scale factor behaves as $a \propto t^p$. Then, the Hubble parameter is given by $H=p/t$
and the solution to Eq.~(\ref{eq_u_t_IR})
is given by using  the Bessel functions as
\beq
u=(mt)^{-\nu} \left( F'~J_\nu(mt) +G'~Y_\nu(mt) \right) \ ,
\label{ugenIR}
\eeq
where
\beq
\nu=\frac{3}{2} p-\frac{1}{2} \ ,
\eeq
and $F'$ and $G'$ are arbitrary constants. 
At early times with $mt < 1$, (i.e., $m < H$),
we can show, 
by using the expansion formula (\ref{bessel_z=0}) of the Bessel function near the origin,
that 
the first and second terms in (\ref{ugenIR}) give those in (\ref{ufrozen}), respectively.
In the MD period, $p=2/3$, $\nu=1/2$, and (\ref{ugenIR}) becomes
\beq
u= (mt)^{-1} \left( F'' \sin (mt) +G'' \cos (mt) \right) \ ,
\label{uMDIR}
\eeq
where $F''$ and $G''$ are arbitrary constants.
Note that the solutions  (\ref{ugenIR}) and (\ref{uMDIR}) are good 
approximations to (\ref{eq_u_t}) 
for low-momentum modes with $k_{\rm phy} \ll m, \sqrt{mH}$, 
either in 
early times $mt <1$ ($m<H$) or in late times $mt>1$ ($m>H$).

As we studied in section~\ref{sec:dark_energy}, 
we are interested in the situation where the wave functions continue 
 to be frozen until the MD period  and then start oscillating. 
Then the coefficients $F''$ and $G''$ in (\ref{uMDIR}) can be determined 
by requiring that $u$ is constant near $t \sim 0$ and the amplitude is given by 
the wave function in the RD period (\ref{uwaveBB}),
or, equivalently, by that in the inflation period (\ref{uBDIRappr}). 
Hence, the wave function in the MD period is given by
\beq
u =  \frac{i H_I}{ \sqrt{2} }
k^{- \frac{3}{2}+\frac{1}{3} \left(\frac{m}{H_I} \right)^2} 
(-\eta_1)^{\frac{1}{3} \left(\frac{m}{H_I} \right)^2}~ 
\frac{\sin(mt)}{mt}  \  .
\label{uMDIR_BD}
\eeq
Using this wave function, the EMT becomes
\beqa
\rho^{\rm IR} &=& \frac{1}{8\pi^2} H_I^2 \frac{1}{t^2}
\left[1-\frac{1}{mt} s + \frac{1}{2(mt)^2}(1-c) \right] N_{\rm eff} \n
&=&\frac{9}{32\pi^2} H_I^2 H^2
\left[1-\frac{3}{2}\frac{H}{m} s + \frac{9}{8} \left( \frac{H}{m} \right)^2 (1-c) \right] N_{\rm eff} \ ,
\label{rhoIR_MD_ETLT}
\eeqa
\beqa
p^{\rm IR} &=& \frac{1}{8\pi^2} H_I^2 \frac{1}{t^2}
\left[c-\frac{1}{mt} s + \frac{1}{2(mt)^2}(1-c) \right] N_{\rm eff} \n
&=&\frac{9}{32\pi^2} H_I^2 H^2
\left[c-\frac{3}{2}\frac{H}{m} s + \frac{9}{8} \left( \frac{H}{m} \right)^2 (1-c) \right] N_{\rm eff} \ ,
\label{pIR_MD_ETLT}
\eeqa
where
\beqa
s &=& \sin (2mt)= \sin \left( \frac{4m}{3H} \right) \ , \n
c &=& \cos (2mt)= \cos \left( \frac{4m}{3H} \right) \ .
\label{scdefinitionMD}
\eeqa
These expressions are valid either in early times $m<H$ ($mt<1$)
or in late times $m>H$ ($mt>1$), since the wave function (\ref{uMDIR}), 
and thus (\ref{uMDIR_BD}), are valid at both times.
Indeed, they give not only the late-time behavior (\ref{IR_MDLT})
but also the early-time behavior (\ref{massMDET}),
as can be seen by expanding the trigonometric functions with respect to  $m/H$. 

We may consider another scenario where the frozen behavior becomes oscillating in the RD period,
as analyzed in section~\ref{sec:EMT}.
In this case, we can connect the wave function (\ref{ugenIR}) in the RD period,
with $p=1/2$ and $\nu=1/4$, to the wave function in the inflation period (\ref{uBDIRappr}), 
and obtain
\beq
u =  \frac{i H_I}{ \sqrt{2} }
k^{- \frac{3}{2}+\frac{1}{3} \left(\frac{m}{H_I} \right)^2} 
(-\eta_1)^{\frac{1}{3} \left(\frac{m}{H_I} \right)^2}
2^\nu \Gamma(\nu+1)
(mt)^{-\nu} J_\nu(mt) \ ,
\label{uk0RD}
\eeq
where we have used the expansion formula (\ref{bessel_z=0}) of the Bessel function near the origin. 
Then,  (\ref{uk0RD}) reproduces the early-time behavior of the EMT (\ref{DEsmallN}).
The late-time behavior is obtained by  using the asymptotic form of the Bessel function
\beq
J_\nu (z) = \sqrt{\frac{2}{\pi z}} \left[\left(1+{\cal O}(z^{-2})\right) \cos \left(z-\frac{2\nu+1}{4}\pi \right) 
+{\cal O}(z^{-1}) \sin \left(z-\frac{2\nu+1}{4}\pi \right)\right]
\eeq
at $|z| \gg 1$. 
It turns out that (\ref{uk0RD}) asymptotes to
the WKB wave function (\ref{uwaveWKB}) with (\ref{WKBqltx}),
which receives higher-order corrections as in (\ref{WKBlm_ho}).
Accordingly, we can reproduce the late-time behavior of the EMT,
(\ref{rholtIR3}) and (\ref{pltIR3}).

%%%%%%%%%%%%%%%%%%%%%%%%%%

\end{document}